\documentclass[12pt, a4paper]{article}
\usepackage{amsfonts, amssymb}
\usepackage{graphicx}
\usepackage{latexsym,amsmath,color,array}

\usepackage{natbib}
\usepackage{enumerate}
\usepackage{bbding}
\usepackage{amssymb}
\usepackage{amsmath}
\usepackage{graphicx}
\usepackage{amsmath}
\usepackage{bbm}
\usepackage{mathtools}
 \usepackage{color}
 \usepackage{array}
 \usepackage{subfigure}
 \usepackage{multirow}
 \usepackage[dvipsnames]{xcolor}
 \usepackage{makecell}
 \usepackage{colortbl}
\usepackage{caption}
\usepackage{booktabs}

\def\1{\mathbb{I}}

\newcounter{thm}[section]
\newcounter{appen}[section]
\newcounter{assum}[section]

\begin{document}

\title{A Multi-parameter regression model for interval censored survival data}
\author{Defen Peng\footnote{University of British Columbia} \hspace{3cm}
Gilbert MacKenzie\footnote{CREST, Ensai; gilbert.mackenzie@ul.ie} \hspace{3cm}
Kevin Burke\footnote{University of Limerick; kevin.burke@ul.ie}  }
\date{}

\maketitle

\begin{abstract}
We develop flexible multi-parameter regression survival models for interval censored survival data arising in longitudinal prospective studies and longitudinal randomised controlled clinical trials. A  multi-parameter Weibull regression survival model, which is wholly parametric, and has non-proportional  hazards, is the main focus of the paper.  We describe the basic model, develop the interval-censored likelihood and extend the model to include gamma frailty and a dispersion model. We evaluate the models by means of a simulation study and a detailed re-analysis  of data from the Signal Tandmobiel$^{\circledR}$ study. The results demonstrate that the multi-parameter regression model with frailty is computationally efficient and provides an excellent fit to the data.
\smallskip


\end{abstract}

\qquad

\newpage

\section{Introduction}
\label{sec:1}
Interval censored survival data can arise in longitudinal epidemiological studies where the response variable $Y(t)$ is binary. Typically, at baseline, $t_0$, patients start in an initial state, e.g.,  for the $i$th patient $Y_i(t_0) =0$ (say) and later, as follow-up proceeds at scheduled inspection times, the event of interest may occur at a time $t_s$ whence, $Y_i(t_s)=1$ , where $t_s>t_0$.  This leads naturally to the use of  ``time to event'' survival modelling in order to determine  the effect of selected risk factors measured at baseline on the time to the event of interest. The interval censoring arises because patients are not monitored continuously but rather  a finite schedule of $m$  follow-up examinations at times $t_k$, $k=1,\dots, m$.  Thus, if, for the $i$th subject, the event occurs between the  $(k-1)$th and the $k$th follow-up examination the binary indicators become $Y_i(t_{s}) =0$   for$ \, s =1,\dots , (k-1)$ and  $Y_i(t_{s}) =1 $ for  $s=k$ and then the time to event is  $t_i \in I_k=(t_{(k-1)},t_k]$.
\par
As a motivating study we consider the Tandmobiel$^{\circledR}$ study which  is a longitudinal prospective oral health study conducted in Flanders (Belgium) from 1996 to 2001. A cohort of 4430 randomly sampled school children who attended the first year of the primary school at the beginning of the study were examined annually (6 times). The response was time to the emergence of the permanent upper left first premolars (tooth 24 in European dental notation). When emergence occurs between annual follow-ups the exact time of emergence is not known and the time to the emergence is said to be interval censored.
\par
Modelling such data presents a variety of challenges. \citet{peng:2013} developed interval censoring methods for parametric models, some of which were non-PH and they compared the use of standard right censored likelihoods based on mid-points with interval censored likelihoods. They showed that the use of mid-points led to artificially precise estimators in PH models when analysing time to loss of vision in a longitudinal trial of Age-Related Macular Degeneration (ARMD) \citep{Hart:2002}.  Survival data arising in longitudinal vision studies, were analysed earlier, by \citet{macKenzie:1999} and later  by \citet{macKenzie:2002}.  See also \citet{fink:1986} for an important early paper in the field, \citet{huang:1997} for a more theoretical review, and the books by \citet{sun:2006} and \citet{bkl:2017} for comprehensive treatments of the subject. Nearly all of these papers employ models where covariates enter through a single parameter. In this paper such models are designated as single parameter regression (SPR) models.
\par
In contrast, the concept of multi-parameter regression (MPR) survival modelling was developed in \citet{kevin:2017}. In MPR survival models the \emph{scale} and \emph{shape} parameters are modelled simultaneously by means of two separate linear predictors: these models  are parametric and intrinsically  more flexible than classical proportional hazards (PH) survival models. In the 2017 paper, MPR models were investigated in the context of right censored survival data from the Northern Ireland Lung Cancer Study \citep{Wilkinson:1995}.
\par
In this paper, we extend  MPR models for interval censored survival data arising in longitudinal studies and introduce a MPR model with gamma frailty and a dispersion model to re-analyse data from the Signal Tandmobiel$^{\circledR}$ study \citep{bogaerts:2002,gomez:2009}.

\section{MPR modelling framework} \label{sec:mpr}
We envisage the class of two-parameter parametric survival models supporting scale and shape parameters. Within that class we model the scale and shape parameters simultaneously by means of two separate linear predictors which may involve the same set, or different sets, of covariates. In this paper  we focus on the Weibull MPR model in order to illustrate the MPR approach to analysing interval-censored data. This model has proved useful in other contexts and has the advantage of directly extending a standard proportional hazards model.

\subsection{Weibull MPR survival model\label{sec:basicmpr}}
The Weibull MPR model of \citet{kevin:2017} can be defined through its hazard function
\begin{equation}
\lambda(t) = \exp(x^T\beta)\exp(z^T\alpha) t^{\exp(z^T\alpha)-1},\label{eq:weib}
\end{equation}
which arises from a basic (no-covariate) Weibull hazard function, $\lambda(t)=\lambda \gamma t^{\gamma-1}$, \citep{lawless:2003,marsholk:2007} allowing both its \emph{scale}, $\lambda>0$, and \emph{shape}, $\gamma>0$, parameters to depend on covariates via the MPR specification
\begin{equation}
\lambda =\exp(x^T\beta) \quad {\rm and} \quad \gamma=\exp(z^T\alpha),\label{eq:mprreg}
\end{equation}
where $x=(1,x_{1},x_{2},\ldots,x_{p})^T$ and $z=(1,z_{1},z_{2},\ldots,z_{q})^T$ are the scale and shape covariate vectors, and $\beta = (\beta_0,\beta_1,\ldots,\beta_p)^T$ and $\alpha = (\alpha_0,\alpha_1,\ldots,\alpha_q)^T$ are the associated regression coefficient vectors. Clearly the hazard function, (\ref{eq:weib}), depends on covariates but we avoid $\lambda(t;x,z)$ for notational convenience; similarly we avoid $\lambda(x)$ and $\gamma(z)$ in (\ref{eq:mprreg}). Note that when $(\alpha_1, \alpha_2,\dots ,\alpha_q)^T=(0,0,\dots,0)^T$ the model reduces to a standard PH regression survival model, i.e., an SPR model. Although the model is non-PH,  when an $\alpha$ coefficient, say $\alpha_j$, is zero, the corresponding covariate is PH,  whence the hazard ratio is $\exp(\beta_j)$ \citep{kevin:2017}. Thus, the Weibull MPR model usefully supports a mix of PH and non-PH effects, including \emph{crossing hazards} and, through the $\alpha$ coefficients, provides covariate-specific tests of proportionality of hazards.

\subsection{Frailty extension}
\label{sec:mprgf}

We extend the Weibull MPR model of \citet{kevin:2017} to incorporate a multiplicative frailty term via the conditional hazard
\begin{equation}
\lambda(t;u)=u\lambda(t)\nonumber,
\end{equation}
where $\lambda(t)$ is defined in (\ref{eq:weib}), and $u$ is the frailty term,  which we will assume follows a gamma distribution with density
\begin{equation}
g(u)=\frac{b^a u^{a-1} \exp(-bu)}{\Gamma(a)}\nonumber,
\end{equation}
where $a=b=1/\phi$, such that $E(U)=1$ and the frailty variance is  $Var(U)=\phi$. This is the classical frailty model in which the random effect, $u$, measures additional  person-specific heterogeneity not accounted for by the covariates. \citep{vaupel:1979}
\par
Since $u$ is an unobserved variable,  the marginal distribution,  obtained by integrating over $u$, has  survivor function given by

\begin{equation}
S_{m}(t)=\{1+\phi\Lambda(t)\}^{-1/\phi},\label{eq:margS}
\end{equation}
where the subscript ``$m$'' indicates marginal, and $\Lambda(t) = \exp(x^T\beta) t^{\exp(z^T\alpha)}$ is the cumulative hazard function associated with (\ref{eq:weib}).
Note that, in the absence of frailty, i.e., $\phi \rightarrow 0$, we have that $S_m(t) \rightarrow \exp\{-\Lambda(t)\}$ restoring the familiar (non-frailty) relationship between a survivor function and its cumulative hazard. See \citet{hougaard:1995,hougaard:2000} and \citet{duchateau:2008} for more details on frailty models.

\subsection{Dispersion model\label{sec:mprdm}}

The MPR frailty model can be extended further with advantage. Usually the frailty variance, $\phi$, is a constant, but  we  can allow the frailty variance to be person-specific via another regression model, i.e.,
\begin{equation}
\phi=\exp(w^T\psi),\label{eq:phireg}
\end{equation}
where $w = (1,w_1,\ldots,w_r)^T$ and $\psi=(\psi_0,\psi_1,\ldots,\psi_r)^T$ are vectors of  covariates and their coefficients respectively. This dispersion model (DM)  allows one to investigate the structure of the dispersion and provides a convenient framework for testing the homogeneity of frailty variances among covariates (e.g., between sexes). In addition, when the frailty variance is unstructured, the model reduces to the Weibull MPR frailty model of Section \ref{sec:mprgf}.
\par
The concept of modelling the structure of the dispersion can be traced back to joint mean-dispersion modelling \citep{smyth:1989,leen:2001, panmac:2003}, but its adoption in the frailty paradigm, in survival analysis, is more recent \citep{Lynch:2014}. Furthermore, the combination of frailty dispersion modelling in combination with an underlying MPR model is novel in the literature.  It will be apparent that the frailty  dispersion model  which introduces a third regression (\ref{eq:phireg}) is an entirely natural development in the MPR paradigm.

\section{Estimation }
\subsection{Likelihood Functions}
In most longitudinal studies the idea of a fixed schedule of follow-ow examinations is too rigid as many subjects fail to respect their exact re-examination appointment dates. Accordingly it is usual to allow the intervals to be person-specific such that  $t_i \in I_{ik}=(t_{i(k-1)},t_{ik}]$.  In general, $t_{ik}$ is close to the scheduled $t_k$ and of course $t_{i0} = 0\,\forall i$. This notation, whilst accurate, is rather cumbersome and it is convenient to  abbreviate it to $I_i=(a_i,b_i]$ in the equations which follow.
\par
 Then a general likelihood for inteval-censoring (IC) data is given by
 \begin{equation*}
L(\theta)= \prod_{i=1}^n
\{S(a_i;\theta)-S(b_i;\theta)\}^{\delta_i} S(t_{ci};\theta)^{1-\delta_i}\label{like:true4}
\end{equation*}
\noindent with subject specific intervals $I_{i}=(a_i,b_i]$, where time $a_i < b_i, \forall \, i$. Here $\delta_i=1$ denotes an interval censored observation and  $\delta_i=0$ denotes a right censored observation with censoring time $t_{ci}$.  In this IC setting, the $i$th subject either  ``fails'' in interval $I_i$,  or is right-censored.   In total, there are $n$ patients of whom $n_c$ are right censored or withdrawn at specific times, leaving $n-n_c$ patients who are interval-censored. Note that the interval-censored subjects play the same role as ``failures'' in the right-censored setting and often right-censoring occurs at times  completely unrelated to the scheduled follow-up examinations, e.g., an early withdrawal from the study.  Thus, this notational setup is advantageous when it is important to distinguish between interval-censored and right-censored observations.  See  \citet{peng:2013} for more details on this approach.
\par
Here, however, we  re-write the likelihood above  as
\begin{equation}
L(\theta)= \prod_{i=1}^n
\{S(a_i;\theta)-S(b_i;\theta)\}\label{like:true3}
\end{equation}
for notational convenience, in which  we define the $i$th right censored observation as lying in the interval $(a_i,b_i] =(t_{ci},\infty]$.  Accordingly, there are now $n$ intervals, with $I_i=(a_i,b_i] \, \forall \, i$. This representation  is the most commonly occurring form in the IC literature and we use it below.
\par
The IC likelihood for the Weibull MPR DM model (i.e., the most general model of Section \ref{sec:mpr}) is
\begin{equation}
L(\theta)= \prod_{i=1}^n
\left[\{1+\phi_i\Lambda(a_i)\}^{-1/\phi_i}-\{1+\phi_i\Lambda(b_i)\}^{-1/\phi_i}\right],\label{like:mprdm}
\end{equation}
where $\theta=(\beta^T,\alpha^T,\psi^T)^T$, $\Lambda(a_i) = \lambda_i a_i^{\gamma_i}$ and $\Lambda(b_i) = \lambda_i b_i^{\gamma_i}$ are the cumulative hazard functions for the $i$th individual evaluated at the end-points of $I_i$ (where, for notational convenience, we avoid expressions such as $\Lambda(a_i;x_i,z_i,w_i)$ and $\Lambda(b_i;x_i,z_i,w_i)$), $\lambda_i = \exp(x_i^T\beta)$ and $\gamma_i = \exp(z_i^T\alpha)$ from (\ref{eq:mprreg}), $\phi_i = \exp(w_i^T\psi)$ from (\ref{eq:phireg}), and $x_i = (1, x_{1i}, \ldots, x_{pi})^T$, $z_i = (1, z_{1i}, \ldots, z_{qi})^T$, and $w_i = (1, w_{1i}, \ldots, w_{ri})^T$ are the scale, shape, and dispersion covariate vectors respectively (which may or may not contain covariates in common).
\par

\subsection{Score Functions}
We now let $\pi_i = S_m(a_i) - S_m(b_i) = \{1+\phi_i\Lambda(a_i)\}^{-1/\phi_i}-\{1+\phi_i\Lambda(b_i)\}^{-1/\phi_i}$ so that the log-likelihood function can be written as $\ell(\theta) = \log L(\theta) = \sum_{i=1}^n \log \pi_i$. The score functions are then given by
\begin{align*}
U(\beta) &=  \sum_{i=1}^n \left[\tfrac{1}{\pi_i}\{S_m(b_i)^{1+\phi_i}\,\omega_{\beta}(b_i) - S_m(a_i)^{1+\phi_i}\,\omega_{\beta}(a_i)\}\,x_i\right], \\
U(\alpha) &= \sum_{i=1}^n \left[\tfrac{1}{\pi_i}\{S_m(b_i)^{1+\phi_i}\,\omega_{\alpha}(b_i) - S_m(a_i)^{1+\phi_i}\,\omega_{\alpha}(a_i)\}\,z_i\right], \\
U(\psi) &= \sum_{i=1}^n \left[\tfrac{1}{\pi_i}\{S_m(b_i)^{1+\phi_i}\,\omega_{\phi}(b_i) - S_m(a_i)^{1+\phi_i}\,\omega_{\phi}(a_i)\}\,w_i\right],
\end{align*}
all of which have a similar functional form, differing only with respect to the $\omega$ ``weight'' functions, which are given by
\begin{align*}
\omega_\beta(a_i) &= \tfrac{\partial}{\partial \beta_0}\Lambda(a_i) = \lambda_i a_i^{\gamma_i}, \\
\omega_\alpha(a_i) &= \tfrac{\partial}{\partial \alpha_0}\Lambda(a_i) = \lambda_i \gamma_i a_i^{\gamma_i} \log a_i, \\
\omega_\psi(a_i) &= \tfrac{\partial}{\partial \psi_0} S_m(a_i) = \Lambda(a_i) + S_m(a_i)^{-\phi_i} \log S_m(a_i),
\end{align*}
with $\omega_\beta(b_i)$, $\omega_\alpha(b_i)$, and $\omega_\psi(b_i)$ analogously defined by replacing $a_i$ with $b_i$.
\par
Although, in the above, we intend that $\Lambda(a_i) = \lambda_i a_i^{\gamma_i}$ for the purpose of the current paper (i.e., that of a Weibull MPR model), we have written the above score functions in a generic form so that $\Lambda(\cdot)$ can be replaced by any cumulative hazard function. 
If the underlying MPR model had another (positive) shape parameter, say, $\rho$, modelled as $\rho = \exp(\tau^Tv)$, then we would gain an additional score function $U(\tau) =  \sum_{i=1}^n \left[(1/\pi_i)\{S_m(b_i)^{1+\phi_i}\,\omega_{\tau}(b_i) - S_m(a_i)^{1+\phi_i}\,\omega_{\tau}(a_i)\}\,v_i\right]$ where $\omega_\tau(a_i) = \partial\Lambda(a_i)/\partial \tau_0$, i.e., this score function has the same structure as that of $U(\beta)$ and $U(\alpha)$, but with a different $\omega$. On the other hand, if the frailty distribution was changed, the $S_m(a_i)^{1+\phi_i}$ factor in all of the score functions would change (and not only through $S_m(\cdot)$ changing), and, of course, the form of $\omega_\psi$ would also change. Thus, although we focus on a Weibull-gamma frailty model, the above is easily adapted to a wide range of MPRDM models for IC data.

\section{Model selection}

For the purpose of selecting among models within the MPRDM class, standard information criteria may be used, namely, the Akaike Information Criterion, AIC $= -2 \log L(\hat \theta) + 2 k$, and the Bayesian Information Criterion, BIC $= - 2\log L(\hat \theta) + (\log n) k$ where $\hat \theta$ is the maximum likelihood estimator and $k = \dim(\theta)$. There are two levels of model selection, both of which can be handled by these information criteria, namely: (a) the overall model type, and (b) the covariate set for each regression component within a given model type.
\par

\subsection{Model types\label{sec:modeltypes}}

The MPRDM modelling framework introduced in Section 2 \ref{sec:mpr} is quite general, containing a range of new and existing regression model types. A natural hierarchy of model types emerges as follows: the underlying model may be PH ($\lambda$ regression) or MPR ($\lambda$ and $\gamma$ regressions), and the frailty component may be absent,  present, or present with a $\phi$ regression component. The six model types are summarised in Table \ref{tab:modeltypes}. Note that models PH and PHF are single parameter regression (SPR) models (only one parameter depends on covariates) while all other types are multi-parameter regression (MPR) models -- in particular, the PHDM model has a PH baseline component (i.e., SPR), but the overall marginal model is MPR since $\phi$ depends on covariates. We have found models without a $\lambda$ regression to be less useful and, so, these are not considered here.

\par
\begin{table}[!htbp]
\small\centering
	\caption{Model types \label{tab:modeltypes}}
	\begin{tabular}{ccccc}
			\toprule
		&&& \multicolumn{2}{c}{~Regression} \\
		Model & Baseline & Frailty & Yes           & No \\
		\hline
		PH    & PH       & No      & $\lambda$            & $\gamma$ \\
		PHF   & PH       & Yes     & $\lambda$            & $\gamma,\phi$ \\
		PHDM  & PH       & Yes     & $\lambda,\phi$       & $\gamma$ \\[0.2cm]
		MPR   & MPR      & No      & $\lambda,\gamma$     &  --- \\
		MPRF  & MPR      & Yes     & $\lambda,\gamma$     & $\phi$ \\
		MPRDM & MPR      & Yes     & $\lambda,\gamma,\phi$ & --- \\
			\bottomrule
	\end{tabular}
	\vskip 2mm
	\captionsetup{width=1.0\textwidth}
	\caption*{\footnotesize ``Model'' is the name of the model; ``Baseline'' is the baseline covariate structure such that ``PH'' is a Proportional Hazards structure where only the scale parameter, $\lambda$, depends on covariates whereas ``MPR'' is a Multi-Parameter Regression structure where the shape parameter, $\gamma$, also depends on covariates (see ``Regression'' columns); ``Frailty'' indicates the presence of a frailty term; ``Regression'' highlights the regression components via the distributional parameters which depend on covariates (``Yes'') and which do not depend on covariates (``No'').}
\end{table}

\subsection{Covariates}

Given a particular model type from Table \ref{tab:modeltypes}, we will generally wish to select from a set of candidate covariates, say $c = (c_0=1, c_1, c_2, \ldots)^T$, to appear in the model (note: $c_0$ is used for the intercept term). In the most general MPRDM model, this amounts to the selection of scale covariates, $x \subset c$, shape covariates, $z \subset c$, and frailty dispersion covariates, $w \subset c$, where the subsets may or may not overlap.
 While the union, $x \cup z \cup w$, is of interest as these covariates affect survival in some way, so too are the $x$, $z$, and $w$ vectors themselves as these characterise the nature of specific covariate effects, e.g, in an MPR model without frailty, $c_j \in z$ implies  that $c_j$ is a non-PH covariate, and, in a frailty dispersion model, $c_j \in w$ indicates that the frailty variance differs in the sub-groups defined by $c_j$.
\par
In general, the basic parameters of survival models (including the frailty variance) are rarely orthogonal, i.e., estimates of these parameters will be correlated. When covariates enter these parameters, this correlation propagates to the regression coefficients. In particular, if the covariate $c_j$ appears in all regression components simultaneously, then the estimates of its corresponding $\beta$, $\alpha$, and $\psi$ coefficients tend to be quite correlated. It is important to emphasize that this correlation does not lead to convergence issues in model fitting, nor does it imply that a covariate must only appear in one regression component within the model. However, it does have implications for variable selection, e.g., individual Wald-based significance tests (which account only for the variance of estimates, and not covariance -- which is important in this context) might render a particular covariate non-significant in all regression coefficients, when, in fact, the overall effect is significant. Covariate selection in MPR models was developed in \citet{kevin:2017} who suggested the use of stepwise procedures in which covariate additions/deletions are carried out for each regression component separately as well as simultaneously, e.g., in an MPR model the covariate $c_j$ could be added to $x$ first (but not to $z$), then to $z$  (but not to $x$), and finally to $x$ and $z$ simultaneously.

\section{Simulation study}
We conducted a simulation study to assess the estimation properties of the Weibull MPR model for interval-censored data. 
Failure times were generated from the Weibull regression model (with or without frailty) with two covariates: $x_1$, a binary covariate where $\Pr(X_1=1)=0.5$ mimicking the treatment effect for example, and $x_2$ a continuous baseline covariate distributed as $N(0,0.5)$.
\par
In addition, we constructed trajectories for each individual in the study by
constructing intervals $I_i =(a_i,b_i]$ such that $a_i = \text{max}(T_i-U_i^{(1)},\, T_i + U_i^{(2)}-c)$ and $b_i = \text{min}(T_i+U_i^{(2)},\, T_i-U_i^{(1)}+c)$, where $U^{(1)}$ and $U^{(2)}$ are independent continuous variables with uniform distribution in the interval $(0, c)$. \citet{zhang:2009} used this approach which, by construction, defines intervals which are non-informative about the survival time distribution, $T$. Furthermore, it can be shown that $E(B_i - A_i)=2c/3$ (proof omitted). In this simulation study, we set $c=(3d/2) E(T)$, i.e., $E(B_i - A_i)= d E(T)$, so that the average inspection length is proportional to the average survival time, $E(T)$, where we use $d \in \{0.1, 0.5\}$.
\par
In the simulations, the proportion of right (random) censoring was controlled by using an exponential distribution where the estimate of the controlling parameter, $\hat\varphi$, was obtained from the ``$J(\cdot)$-function'' approach of \citet{peng:2013}. Suppose the independent censoring times follow an exponential distribution with density $g(t; \varphi)$. Let
$$J(\varphi) = \Big[\int_0^{\infty} S(t; \theta)g(t; \varphi)dt - p\Big]^2$$
where $S(t;\theta)$ is the survival function
and  $p$ is the  censoring proportion  required. Then, $\hat\varphi = \text{arg}\, \text{min} [J(\varphi)]$ ensures that, on average, proportion of censored individuals in each simulation equals $p$. We set $p\in\{0, 0.3\}$ in this simulation study. The entire simulation was conducted in the R software package. \citep{rteam:2018}
\par
Three sample sizes were used $n \in \{200, 500,1000\}$. Each scenario was replicated 5000 times, and, for each replicate, the model was fitted using the likelihood function given in (\ref{like:mprdm}). Note that the true coefficient values were set as $\beta_0=2.0, \beta_1=0.5, \beta_2=0.3; \alpha_0=2.0, \alpha_1=0.25, \alpha_2=-0.1; \phi=0.5, \phi_1=0.15, \phi_2=-0.2$.

\subsection{Results}
Table \ref{tab:simnonfrailty} shows the median estimate, standard error, and average relative percentage bias ($= 100 \times \frac{\hat\theta-\theta}{\theta}$) for each parameter arising from our simulation study. We see that the estimators from the MPR IC likelihood have very little bias, and that both the bias and standard error tend to reduce with increasing sample size. As expected, larger right-censoring and inspection tend to reduce performance, but, even in the worst case of 30\% right-censoring and average inspection length of 0.5 $E(T)$, the results are quite good.
\par
Table \ref{tab:simfrailty} and Table \ref{tab:simfrailtydm} show the results of the simulation from the Weibull MPR with frailty, $\phi=0.5$ and Weibull MPR with dispersion model, $\phi=0.5, \phi_1=0.15, \phi_2=-0.2$, respectively. For the smallest sample size ($n=200$) the frailty estimates can be somewhat biased (and even for $n=500$ for the more complicated dispersion model case), but the bias is much reduced for $n=1000$. Of course, we expect that more complicated models tend to require larger sample sizes, but, overall, our simulation suggests that the model parameters are recoverable for reasonable sample sizes.

\begin{table}[!htbp]
	\small\centering
	\caption{Simulation: ML Estimates for the MPR Weibull model with various sample sizes and censoring rates -- without frailty$^\dag$\label{tab:simnonfrailty}}
	\footnotesize
	\begin{tabular}{lccclccc}
		\toprule
		Estimates & $\hat\beta$ & $SE$  & $\%bias$ & & $\hat\alpha$ & $SE$  & $\%bias$ \\[0.5ex]
		\hline
		& & & & & & & \\
		$n=200$ & & & \multicolumn{3}{c}{$p=0, d=0.1$} &  & \\
		Intercept         & 2.04  & 0.11   & 2.55  &     & 2.04   & 0.08    &2.94 \\[-0.5ex]
		$x_1$             & 0.51  & 0.16   & 1.15  &     & 0.25   & 0.11    &-0.71\\[-0.5ex]
		$x_2$             & 0.30  & 0.16   & 0.82  &     & -0.11  & 0.11    &5.08 \\[-0.5ex]
		
		$n=500$ &         &  &           & &     &   & \\
		Intercept         & 2.01   & 0.06        &0.62 &     & 2.01    & 0.05    &1.05\\[-0.5ex]
		$x_1$             & 0.51   & 0.10        & 1.51&     & 0.25    & 0.07    &1.05\\[-0.5ex]
		$x_2$             & 0.30   & 0.10        &0.27&      & -0.10   & 0.07    &3.22\\[-0.5ex]
		
		$n=1000$ &         &  &           & &     &   & \\
		Intercept         & 2.00   & 0.05        &0.32&      & 2.01    & 0.04    &0.73 \\[-0.5ex]
		$x_1$             & 0.50   & 0.07        &0.84&      & 0.25    & 0.05    &0.10 \\[-0.5ex]
		$x_2$             & 0.30   & 0.07        &0.47 &     & -0.10   & 0.05    &0.14\\[-0.5ex]
		&   & &  & &   &   &    \\[-0.5ex]
		$n=200$ & & & \multicolumn{3}{c}{$p=0.3, d=0.1$} &  & \\
		Intercept         & 2.04  & 0.13        &2.84 &     & 2.04   & 0.09    &3.14 \\[-0.5ex]
		$x_1$             & 0.51  & 0.20        &1.32 &     & 0.25   & 0.13    &0.31\\[-0.5ex]
		$x_2$             & 0.30  & 0.20        &1.00 &     & -0.10  & 0.13    &1.86 \\[-0.5ex]
		
		$n=500$ &         &  &           & &     &   & \\
		Intercept         & 2.01  & 0.08        &0.97 &     & 2.02   & 0.06    &1.33 \\[-0.5ex]
		$x_1$             & 0.51  & 0.12        &1.21 &     & 0.25   & 0.08    &0.66 \\[-0.5ex]
		$x_2$             & 0.30  & 0.13        &0.86 &    & -0.10   & 0.08    &0.96 \\[-0.5ex]
		
		$n=1000$ &         &  &           & &     &   & \\
		Intercept         & 2.01  & 0.06        &0.51 &      & 2.01   & 0.04    &0.78 \\[-0.5ex]
		$x_1$             & 0.50  & 0.09        & 0.44 &    & 0.25   & 0.06    &0.05 \\[-0.5ex]
		$x_2$             & 0.30  & 0.09        &0.03 &      & -0.10  & 0.06    &0.79 \\[-0.5ex]
		&   & &  & &   &   &    \\[-0.5ex]
		$n=200$ & & & \multicolumn{3}{c}{$p=0, d=0.5$} &  & \\
		Intercept         & 2.03  & 0.11        &1.99 &      & 2.03   & 0.09    &2.20 \\[-0.5ex]
		$x_1$             & 0.51  & 0.17        &1.08 &      & 0.25   & 0.12    &0.48 \\[-0.5ex]
		$x_2$             & 0.30  & 0.17        &1.29 &      & -0.10  & 0.13    &-1.10 \\[-0.5ex]
		
		$n=500$ &         &  &           & &     &   & \\
		Intercept         & 2.01  & 0.07        &1.02&       & 2.02   & 0.05    &1.19\\[-0.5ex]
		$x_1$             & 0.50  & 0.11        &0.58 &      & 0.25   & 0.08    &0.45 \\[-0.5ex]
		$x_2$             & 0.30  & 0.11        &1.29 &     & -0.10  & 0.08    &2.17 \\[-0.5ex]
		
		$n=1000$ &         &  &           & &     &   & \\
		Intercept         & 2.01  & 0.05        &0.41&      & 2.01   & 0.04    &0.59\\[-0.5ex]
		$x_1$             & 0.50  & 0.08        & 0.71&     & 0.25   & 0.06    &0.57 \\[-0.5ex]
		$x_2$             & 0.30  & 0.07        &-0.82 &     & -0.10  & 0.06    &0.01 \\[-0.5ex]
		&   & &  & &   &   &    \\[-0.5ex]
	$n=200$ & & & \multicolumn{3}{c}{$p=0.3, d=0.5$} &  & \\
	Intercept         & 2.03  & 0.13        &2.21 &      & 2.05   & 0.10    &3.79 \\[-0.5ex]
	$x_1$             & 0.51  & 0.22        &2.83&      & 0.25   & 0.14    &1.16 \\[-0.5ex]
	$x_2$             & 0.30  & 0.21        &0.14 &      & -0.10  & 0.14    &0.32 \\[-0.5ex]
	
	$n=500$ &         &  &           & &     &   & \\
	Intercept         & 2.00  & 0.08        &0.27&       & 2.03   & 0.06    &2.27\\[-0.5ex]
	$x_1$             & 0.51  & 0.13        &1.32 &      & 0.25   & 0.09    &1.32 \\[-0.5ex]
	$x_2$             & 0.30  & 0.13        &0.94 &     & -0.10  & 0.09    &-2.83 \\[-0.5ex]
	
	$n=1000$ &         &  &           & &     &   & \\
	Intercept         & 2.00  & 0.06        &0.09&      & 2.02   & 0.04    &1.67\\[-0.5ex]
	$x_1$             & 0.50  & 0.09        & 0.11&     & 0.25   & 0.06    &0.86 \\[-0.5ex]
	$x_2$             & 0.30  & 0.09        &0.33 &     & -0.10  & 0.06    &-1.55 \\[-0.5ex]
	&   & &  & &   &   &    \\[-0.5ex]
		\bottomrule
	\end{tabular}
	\vskip 1mm
	\captionsetup{width=1.0\textwidth}
	\caption*{\footnotesize$^\dag$ The true parameters: $\lambda=2.0, \beta_1=0.5, \beta_2=0.3; \alpha=2.0, \alpha_1=0.25, \alpha_2=-0.1$.\qquad\qquad}
\end{table}

\begin{table}[!htbp]
\small\centering
	\caption{Simulation: ML Estimates for the MPR Weibull model with various numbers of sample sizes and censoring rates -- with frailty variance $\phi=0.5 $$^\dag$  \label{tab:simfrailty}}
	\footnotesize
	\begin{tabular}{lccclccclccc}
		\toprule
		Estimates & $\hat\beta$ & $SE$  & $\%bias$ & & $\hat\alpha$ & $SE$  & $\%bias$ & & $\hat\phi$ & $SE$  & $\%bias$\\[0.5ex]
		\hline
		& & & & & & & \\
		$n=200$ & & &  & &  \multicolumn{3}{c}{$p=0, d=0.1$} & & & & \\
		Intercept         & 1.99 & 0.22 &-0.80&     & 2.02 &0.12  &1.49& & 0.48 &0.40  &5.87 \\[-0.5ex]
		$x_1$             & 0.51 &0.23  &1.08&      & 0.25 &0.12  &1.12& & -&-&- \\[-0.5ex]
		$x_2$             &0.30 &0.22 &-0.38&      & -0.10 &0.12  &2.47& & -&-&- \\[-0.5ex]
	
		$n=500$ &         &  &           & &     &   & \\
		Intercept         &2.00 &0.14  &0.28 &     & 2.01 &0.07  &1.03 & & 0.50 &0.24  &0.81\\[-0.5ex]
		$x_1$             & 0.50 &0.14  &0.26 &     & 0.25 &0.07  &0.44 & & -&-&-\\[-0.5ex]
		$x_2$             & 0.30 &0.14  &0.99 &     & -0.10 &0.07 &-0.46 & & -&-&-\\[-0.5ex]
		
		$n=1000$ &         &  &           & &     &   & \\
		Intercept         & 2.00 &0.10  &0.00&      & 2.00 &0.05  &0.35 & & 0.50 &0.17  &0.95\\[-0.5ex]
		$x_1$             & 0.50 &0.10  &0.13 &     & 0.25 &0.05  &0.37 & & -&-&-\\[-0.5ex]
		$x_2$             & 0.30 &0.10  &1.06&    & -0.10 &0.05 &-0.11 & & -&-&-\\[-0.5ex]
		
		& & & & & & & \\
		$n=200$ & & &  & &  \multicolumn{3}{c}{$p=0.3, d=0.1$} & & & & \\
		Intercept         & 2.02 &0.28  &1.44 &      & 2.03 &0.14  &2.26& & 0.48 &0.53  &4.53\\[-0.5ex]
		$x_1$             &0.51 &0.27  &2.53&      & 0.26 &0.14  &2.12 & & -&-&-\\[-0.5ex]
		$x_2$             & 0.30 &0.26  &1.06&      &-0.10 &0.13 &-4.50 & & -&-&-\\[-0.5ex]
		
		$n=500$ &         &  &           & &     &   & \\
		Intercept         & 2.01 &0.18  &0.76 &      & 2.02 &0.09  &1.24& & 0.50 &0.32  &0.76\\[-0.5ex]
		$x_1$             &0.50 &0.17  &0.68&      & 0.25 &0.09  &1.23 & & -&-&-\\[-0.5ex]
		$x_2$             & 0.30 &0.18 &-1.43&      &-0.10 &0.09  &3.46 & & -&-&-\\[-0.5ex]
		
		$n=1000$ &         &  &           & &     &   & \\
		Intercept         & 2.01 &0.12  &0.66 &      & 2.01 &0.06  &0.97 & & 0.50 &0.23 &-0.18\\[-0.5ex]
		$x_1$             & 0.51 &0.12  &1.45&       & 0.25 &0.06  &0.32 & & -&-&-\\[-0.5ex]
		$x_2$             & 0.30 &0.13  &1.39 &     & -0.10 &0.06  &0.59 & & -&-&-\\[-0.5ex]
		
		& & & & & & & \\
		$n=200$ & & &  & &  \multicolumn{3}{c}{$p=0, d=0.5$} & & & & \\
		Intercept         & 1.98 &0.25 &-1.55&      & 2.01 &0.14  &0.88 & & 0.48 &0.45  &7.05\\[-0.5ex]
		$x_1$             & 0.50 &0.24 &-0.15&     & 0.25 &0.13  &1.80 & & -&-&-\\[-0.5ex]
		$x_2$             &0.30 &0.22 &-0.06 &    & -0.10 &0.13  &4.62 & & -&-&-\\[-0.5ex]
		
		$n=500$ &         &  &           & &     &   & \\
		Intercept         & 1.99 &0.16 &-0.95 &       & 2.00 &0.09  &0.19 & & 0.49 &0.27  &3.69\\[-0.5ex]
		$x_1$             & 0.50 &0.15  &0.15&       & 0.25 &0.08 &-0.24 & & -&-&-\\[-0.5ex]
		$x_2$             &0.30 &0.15 &-0.79&      & -0.10 &0.08  &0.46 & & -&-&-\\[-0.5ex]
		
		$n=1000$ &         &  &           & &     &   & \\
		Intercept         & 2.00 &0.11 &-0.16 &     & 2.00 &0.06  &0.02& & 0.49 &0.19  &1.68\\[-0.5ex]
		$x_1$             & 0.50 &0.11  &0.52&      & 0.25 &0.06  &0.70 & & -&-&-\\[-0.5ex]
		$x_2$             & 0.30 &0.10  &0.39 &     & -0.10 &0.06 &-0.33 & & -&-&-\\[-0.5ex]
		
		& & & & & & & \\
		$n=200$ & & &  & &  \multicolumn{3}{c}{$p=0.3, d=0.5$} & & & & \\
		Intercept         & 2.01 &0.32  &0.41&       & 2.05 &0.16  &3.65 & & 0.48 &0.60  &5.41\\[-0.5ex]
		$x_1$             &  0.50 &0.29  &0.96&       & 0.25 &0.15 &-0.71 & & -&-&-\\[-0.5ex]
		$x_2$             & 0.31 &0.29  &3.71&      & -0.10 &0.15  &0.55 & & -&-&-\\[-0.5ex]
		
		$n=500$ &         &  &           & &     &   & \\
		Intercept         & 1.99 &0.19 &-0.72&       & 2.03 &0.10  &2.29& & 0.50 &0.36  &0.66\\[-0.5ex]
		$x_1$             &  0.50 &0.18  &0.60&       & 0.25 &0.09  &0.11 & & -&-&-\\[-0.5ex]
		$x_2$             & 0.30 &0.17  &1.55&      & -0.10 &0.09 &-4.23 & & -&-&-\\[-0.5ex]
	
		$n=1000$ &         &  &           & &     &   & \\
		Intercept         & 2.00 &0.13  &0.00 &     & 2.02 &0.07  &1.71 & & 0.50 &0.25 &-0.42\\[-0.5ex]
		$x_1$             & 0.50 &0.12  &0.49&      & 0.25 &0.07  &0.94 & & -&-&-\\[-0.5ex]
		$x_2$             & 0.30 &0.12  &1.52 &     & -0.10 &0.07 &-1.61 & & -&-&-\\[-0.5ex]
		
		&   & &  & &   &   &    \\[-0.5ex]
		\bottomrule
	\end{tabular}
	\vskip 1mm
	\captionsetup{width=1.0\textwidth}
	\caption*{\footnotesize $^\dag$ The true parameters: $\lambda=2.0, \beta_1=0.5, \beta_2=0.3; \alpha=2.0, \alpha_1=0.25, \alpha_2=-0.1; \phi=0.5$.\qquad\qquad}
\end{table}

\begin{table}[!htbp]
	\small\centering
	\caption{Simulation: ML Estimates for the MPR Weibull model with various numbers of sample sizes and censoring rates -- with frailty dispersion \label{tab:simfrailtydm}}
	\footnotesize
	\begin{tabular}{lccclccclccc}
		\toprule
		Estimates & $\hat\beta$ & $SE$  & $\%bias$ & & $\hat\alpha$ & $SE$  & $\%bias$ & & $\hat\phi$ & $SE$  & $\%bias$\\[0.5ex]
		\hline
		& & & & & & & \\
		$n=200$ & & &  & &  \multicolumn{3}{c}{$p=0, d=0.1$} & & & & \\
		Intercept         & 2.01 &0.29   &0.57&     & 2.03 &0.14   &2.27&   & 0.46 &0.57  &12.54 \\[-0.5ex]
		$x_1$             & 0.48 &0.44   &-3.53&    & 0.24 &0.20   &-2.51&  & 0.15 &0.78  &-0.67\\[-0.5ex]
		$x_2$             & 0.32 &0.32   &5.77&     & -0.09 &0.15  &-6.89&  & -0.18 &0.56 &-11.71 \\[-0.5ex]
		
		$n=500$ &         &  &           & &     &   & \\
		Intercept         & 2.00 &0.17 &-0.11 &     & 2.01 &0.09  &0.62 & & 0.48 &0.33  &5.47\\[-0.5ex]
		$x_1$             & 0.51 &0.28  &2.04&      & 0.25 &0.13  &1.44 & & 0.17& 0.47 &10.98\\[-0.5ex]
		$x_2$             & 0.30 &0.24  &1.39 &     & -0.10 &0.11  &0.88 & & -0.20 &0.41 &-0.68\\[-0.5ex]
		
		$n=1000$ &         &  &           & &     &   & \\
		Intercept         & 2.00 &0.12  &0.01&      & 2.00 &0.06  &0.31  & & 0.49 &0.24  &3.07\\[-0.5ex]
		$x_1$             & 0.50 &0.20  &0.99 &     & 0.25 &0.09  &0.64  & & 0.15 &0.32  &2.68\\[-0.5ex]
		$x_2$             & 0.31 &0.18  &2.12&      & -0.10 &0.08 &-1.31 & & -0.20 &0.29 &-2.08\\[-0.5ex]
		
		& & & & & & & \\
		$n=200$ & & &  & &  \multicolumn{3}{c}{$p=0.3, d=0.1$} & & & & \\
		Intercept         & 2.11 &0.35   &7.59 &     & 2.06 &0.16   &4.25  & &0.48 &0.72   &5.93\\[-0.5ex]
		$x_1$             & 0.50 &0.56   &0.81&      & 0.25 &0.22   &1.51  & & 0.13 &1.02 &-11.36\\[-0.5ex]
		$x_2$             & 0.31 &0.43   &4.02&      &-0.10 &0.18   &-2.31 & & -0.19 &0.75  &-7.38\\[-0.5ex]
		
		$n=500$ &         &  &           & &     &   & \\
		Intercept         & 2.01 &0.21  &0.50 &     & 2.01  &0.10  &1.04  & & 0.47 &0.47  &8.34\\[-0.5ex]
		$x_1$             & 0.52 &0.34  &3.76&      &  0.25 &0.14  &0.40  & & 0.19 &0.65 &24.02\\[-0.5ex]
		$x_2$             & 0.31 &0.26  &3.92&      & -0.10 &0.11  &-2.16 & & -0.18 &0.49 &-8.27\\[-0.5ex]
		
		$n=1000$ &         &  &           & &     &   & \\
		Intercept         & 2.00 &0.16  &0.17 &      & 2.01 &0.07  &0.45   & & 0.49 &0.34  &4.34\\[-0.5ex]
		$x_1$             & 0.51 &0.24  &2.36&       & 0.26 &0.10  &2.18   & & 0.17 &0.44  &13.61\\[-0.5ex]
		$x_2$             & 0.31 &0.20  &2.51 &     & -0.10 &0.09  &-0.95  & &-0.20 &0.37  &0.24\\[-0.5ex]
		
		& & & & & & & \\
		$n=200$ & & &  & &  \multicolumn{3}{c}{$p=0, d=0.5$} & & & & \\
		Intercept         & 1.99 &0.32  &-0.69&      & 2.02 &0.18   & 1.11  & & 0.44 &0.66  &19.10\\[-0.5ex]
		$x_1$             & 0.49 &0.50  &-2.86&      & 0.25 &0.24   &-0.94  & & 0.19 &0.90  &29.52\\[-0.5ex]
		$x_2$             & 0.31 &0.32  &1.84 &      & -0.10 &0.16  &-2.32  & & -0.18 &0.56 &-11.54\\[-0.5ex]
		
		$n=500$ &         &  &           & &     &   & \\
		Intercept         & 1.98 &0.19  &-1.66 &       & 2.00  &0.11  &0.10  & & 0.47 &0.38  &8.38\\[-0.5ex]
		$x_1$             & 0.50 &0.31  &-0.10&        & 0.25  &0.15  &0.88  & & 0.17 &0.52 &13.91\\[-0.5ex]
		$x_2$             & 0.31 &0.23  &4.95&         & -0.09 &0.11  &-5.28 & & -0.19 &0.39 &-4.39\\[-0.5ex]
		
		$n=1000$ &         &  &           & &     &   & \\
		Intercept         & 2.00 &0.14  &-0.33 &     & 2.00 &0.07  &0.07  & & 0.49 &0.26  &4.14\\[-0.5ex]
		$x_1$             & 0.50 &0.22  &-0.81&      & 0.25 &0.11  &1.43  & & 0.16 &0.36  &3.93\\[-0.5ex]
		$x_2$             & 0.30 &0.19  &1.49 &      & -0.10 &0.09 &-1.22 & & -0.20 &0.31 &-1.01\\[-0.5ex]
		
		& & & & & & & \\
		$n=200$ & & &  & &  \multicolumn{3}{c}{$p=0.3, d=0.5$} & & & & \\
		Intercept         & 2.12 &0.39  &8.46&       & 2.11  &0.19  &7.63   & & 0.49 &0.79   &3.67\\[-0.5ex]
		$x_1$             & 0.50 &0.62  &0.90&       & 0.24  &0.26  &-2.37  & & 0.13 &1.09 &-13.02\\[-0.5ex]
		$x_2$             & 0.34 &0.45  &14.57&      & -0.09 &0.20  &-12.54 & & -0.15 &0.75 &-24.64\\[-0.5ex]
		
		$n=500$ &         &  &           & &     &   & \\
		Intercept         & 2.01 &0.24  &0.88&       & 2.04 &0.12   &2.77  & & 0.48 &0.52   &7.28\\[-0.5ex]
		$x_1$             & 0.52 &0.38  &3.62&       & 0.25 &0.17   &1.39  & & 0.17 &0.70  &12.32\\[-0.5ex]
		$x_2$             & 0.31 &0.28  &2.48&       & -0.10 &0.13  &-4.14 & & -0.18 &0.50 &-11.03\\[-0.5ex]
		
		$n=1000$ &         &  &           & &     &   & \\
		Intercept         & 2.01 &0.17 &10.65 &     & 2.03 &0.08  &2.48  & & 0.49 &0.36  &1.58\\[-0.5ex]
		$x_1$             & 0.50 &0.27 &-0.80&      & 0.25 &0.12  &-0.42 & & 0.14 &0.48 &-3.85\\[-0.5ex]
		$x_2$             & 0.31 &0.22 &2.51 &      &-0.09 &0.10  &-6.28 & & -0.20 &0.39 &-0.80\\[-0.5ex]
		
		&   & &  & &   &   &    \\[-0.5ex]
		\bottomrule
	\end{tabular}
	\vskip 1mm
	\captionsetup{width=1.0\textwidth}
	\caption*{\footnotesize $^\dag$ The true parameters: $\lambda=2.0, \beta_1=0.5, \beta_2=0.3; \alpha=2.0, \alpha_1=0.25, \alpha_2=-0.1; \phi=0.5, \phi_1=0.15, \phi_2=-0.2$.\qquad\qquad}
\end{table}

\section{Data analysis }
We analysed a subset of data from the  Signal Tandmobiel$^{\circledR}$study.  Following other authors (e.g., \citet{gomez:2009}), time to emergence, $T$, was measured as ``child's age minus five years'' (i.e., age $-$ 5)  since emergence of pre-molars does not occur before age five; more generally, of course, this threshold parameter could be estimated. Two covariates were analysed, namely, sex, where 0 = \text{boy}\, (52\%) and 1 = \text{girl}\, (48\%), and dmf, where 0 (57\%) indicates that the primary predecessor tooth was sound and 1 (43\%) indicates that it was decayed, missing due to caries, or filled. \citet{gomez:2009} excluded 44 (1\%) school children in whom the dmf status was unknown thus leaving 4386 children for our analysis. It should be noted that a more extensive data set from the Signal Tandmobiel$^{\circledR}$ study has been analysed by \citet{bogaerts:2002}, \citet{kl:2005} and \citet{kl:2009}.
\par
We investigated four covariate structures: (I) sex only, (II) dmf only, (III) sex and dmf together, and (IV) sex and dmf together along with their interaction term. Each of these covariate structures were included in the six different model types designated by: PH, PHF, PHDM, MPR, MPRF, MPRDM (see Table\ref{tab:modeltypes}).  Once the best-fitting model type is identified (using  the mean AIC averaged over covariate structures as a guide), covariate selection proceeds within the best type, and the final  model selected, may, if MPR, have different covariates in the regressions.
\par
Thus, there are 24 initial models in total defined by the combination of four covariate structures in each of six model types. These models were fitted to the data using a specially written \texttt{R} programme \citep{rteam:2018} which called  the routine \verb!nlm! to maximize the likelihood function (\ref{like:mprdm}) and compute the observed information matrix. Furthermore, we  computed the non-parametric maximum likelihood estimator of the survivor function \citep{turnbull:1976} using the \texttt{R} package \texttt{interval} \citep{fay:2010}.


\subsection{Results}

A summary of the 24 initial models fitted to the data is given in Table \ref{tab:fit}. Firstly, looking at the four regression structures (I) - (IV), we see that, for all model types, the dmf models, (II), have lower AICs and BICs than the sex models, (I), suggesting that the status of the primary predecessor tooth (sound versus decayed/missing/filled) has a greater effect than the sex of the child. This is not to say that sex is unimportant, as the simultaneous inclusion of both sex and dmf, (III), yields a greater reduction in AIC and BIC relative to the two single-factor models. On the other hand, interestingly, the addition of the interaction term, (IV), reduces the AICs, but increases the BICs, suggesting that there may be a weak interaction effect.

\begin{table}[!htbp]
	\footnotesize\centering
	\caption{Summary of initial models  fitted \label{tab:fit}}
		\begin{tabular}{lccc@{}rcccc}
		\toprule
			&&&&&&&&\\[-0.3cm]
			Model & Covariates & $\ell(\hat\theta)$ & $\dim(\theta)$ && AIC & BIC &  dAIC &  dBIC \\
			\hline
			PH(I)   & sex                & -5562.1 &     3 && 11130.1 & 11149.3 & 180.0 & 147.3 \\       
			PH(II)  & dmf                & -5559.2 &     3 && 11124.5 & 11143.7 & 174.4 & 141.7 \\       
			PH(III) & sex$\,+\,$dmf      & -5523.9 &     4 && 11055.7 & 11081.3 & 105.6 &  79.3 \\       
			PH(IV)  & sex$\,\times\,$dmf & -5520.2 &     5 && 11050.3 & 11082.3 & 100.2 &  80.3 \\       
			\cline{6-9}
			&&&&&&&&\\[-0.3cm]                                                                %
			&&&& mean: & 11090.2 & 11114.1 & 140.0 & 112.2 \\[0.4cm]                                       %
			PHF(I)   & sex                & -5540.9 &     4 && 11089.7 & 11115.3 & 139.6 & 113.3 \\       
			PHF(II)  & dmf                & -5526.6 &     4 && 11061.2 & 11086.7 & 111.0 &  84.8 \\       
			PHF(III) & sex$\,+\,$dmf      & -5488.2 &     5 && 10986.4 & 11018.3 &  36.3 &  16.4 \\       
			PHF(IV)  & sex$\,\times\,$dmf & -5485.1 &     6 && 10982.2 & 11020.5 &  32.0 &  18.5 \\       
			\cline{6-9}
			&&&&&&&&\\[-0.3cm]                                                                %
			&&&& mean: & 11029.9 & 11060.2 & 79.7 & 58.3 \\[0.4cm]                                       %
			PHDM(I)   & sex                & -5540.8 &     5 && 11091.7 & 11123.6 & 141.5 & 121.6 \\       
			PHDM(II)  & dmf                & -5516.3 &     5 && 11042.5 & 11074.5 &  92.4 &  72.5 \\       
			PHDM(III) & sex$\,+\,$dmf      & -5475.2 &     7 && 10964.4 & 11009.1 &  14.3 &   7.1 \\       
			PHDM(IV)  & sex$\,\times\,$dmf & -5472.6 &     9 && 10963.3 & 11020.8 &  13.2 &  18.8 \\       
			\cline{6-9}
			&&&&&&&&\\[-0.3cm]                                                                %
			&&&& mean: & 11015.5 & 11057.0 & 65.3 & 55.0 \\[0.4cm]                                       %
			MPR(I)   & sex                & -5560.8 &     4 && 11129.7 & 11155.2 & 179.6 & 153.3 \\       
			MPR(II)  & dmf                & -5538.3 &     4 && 11084.6 & 11110.2 & 134.5 & 108.2 \\       
			MPR(III) & sex$\,+\,$dmf      & -5501.7 &     6 && 11015.4 & 11053.7 &  65.3 &  51.8 \\       
			MPR(IV)  & sex$\,\times\,$dmf & -5493.7 &     8 && 11003.4 & 11054.4 &  53.2 &  52.5 \\       
			\cline{6-9}
			&&&&&&&&\\[-0.3cm]                                                                %
			&&&& mean: & 11058.3 & 11093.4 & 108.1 & 91.4 \\[0.4cm]                                       %
			MPRF(I)   & sex                & -5540.7 &     5 && 11091.4 & 11123.4 & 141.3 & 121.4 \\       
			MPRF(II)  & dmf                & -5511.3 &     5 && 11032.6 & 11064.5 &  82.5 &  62.6 \\       
			MPRF(III) & sex$\,+\,$dmf      & -5471.6 &     7 && 10957.2 & 11001.9 &   7.1 &   0.0 \\       
			MPRF(IV)  & sex$\,\times\,$dmf & -5466.1 &     9 && 10950.1 & 11007.6 &   0.0 &   5.7 \\       
			\cline{6-9}
			&&&&&&&&\\[-0.3cm]                                                                %
			&&&& mean: & 11007.8 & 11049.4 & 57.7 & 47.4 \\[0.4cm]                                       %
			MPRDM(I)   & sex                & -5540.7 &     6 && 11093.3 & 11131.7 & 143.2 & 129.7 \\       
			MPRDM(II)  & dmf                & -5511.2 &     6 && 11034.4 & 11072.7 &  84.3 &  70.8 \\       
			MPRDM(III) & sex$\,+\,$dmf      & -5469.8 &     9 && 10957.6 & 11015.1 &   7.5 &  13.2 \\       
			MPRDM(IV)  & sex$\,\times\,$dmf & -5465.6 &    12 && 10955.2 & 11031.9 &   5.1 &  29.9 \\       
			\cline{6-9}
			&&&&&&&&\\[-0.3cm]                                                                %
			&&&& mean: & 11010.2 & 11062.8 & 60.0 & 60.9 \\[0.4cm]
			\bottomrule
		\end{tabular}
	\vskip 2mm
	\captionsetup{width=1.0\textwidth}
	\caption*{\footnotesize
		``sex'' and ``dmf'' indicate a single-factor model in which one of sex or dmf appears; ``sex$+$dmf'' indicates a model with both sex and dmf; ``sex$\times$dmf'' indicates a models with both sex and dmf along with the interaction between these two; note that all models contain an intercept terms in the scale and shape; $\ell(\hat\theta)$ is the log-likelihood value; $\dim(\theta)$ is the number of parameters in the model; ``AIC'' and ``BIC'' are the Akaike Information Criteria and Bayesian Information Criteria respectively;  $\text{dAIC} = \text{AIC} - \min(\text{AIC})$ where $\min(\text{AIC})$ represents the lowest AIC among the models shown in this table which is that of model MPRF(IV) whose AIC is $10950.1$; $\text{dBIC} = \text{BIC} - \min(\text{BIC})$ where $\min(\text{BIC})$ corresponds to model MPRF(III) whose BIC is 11001.9. Note that, for each of the six model types, the mean AIC, BIC, dAIC, and dBIC values are shown to facilitate quick comparison of the basic model types.}
\end{table}

\begin{figure}[htbp]
	\begin{center}
		\begin{tabular}{@{\hspace{-0.2cm}}l@{\quad}l}
			\qquad{\bf (a)}\quad PH(IV) -- Boys & \qquad{\bf (b)}\quad PH(IV) -- Girls \\
			\includegraphics[width=0.5\textwidth, trim = {0cm 0cm 0cm 2.09cm}, clip]{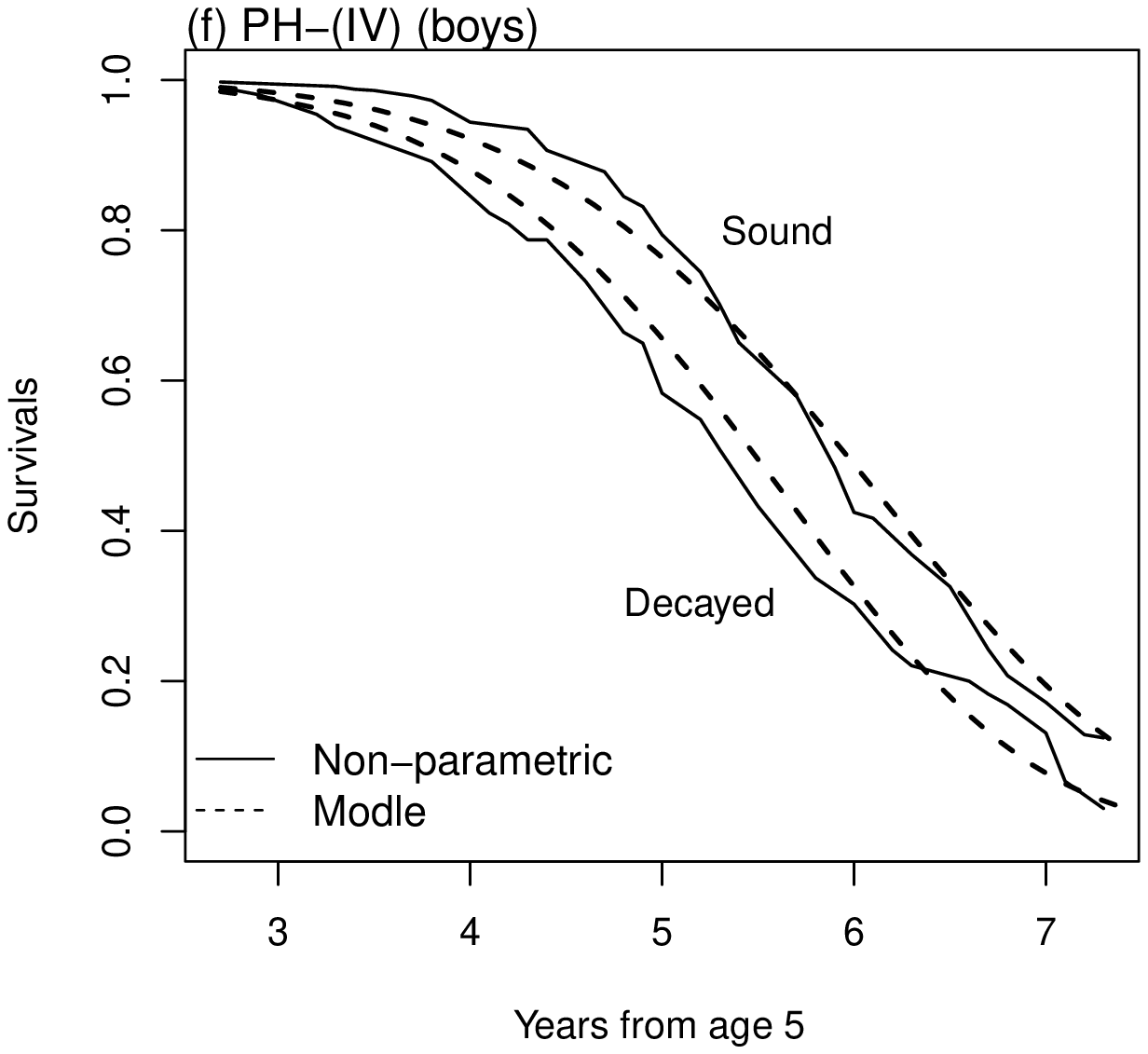} & \includegraphics[width=0.5\textwidth, trim = {0cm 0cm 0cm 2.09cm}, clip]{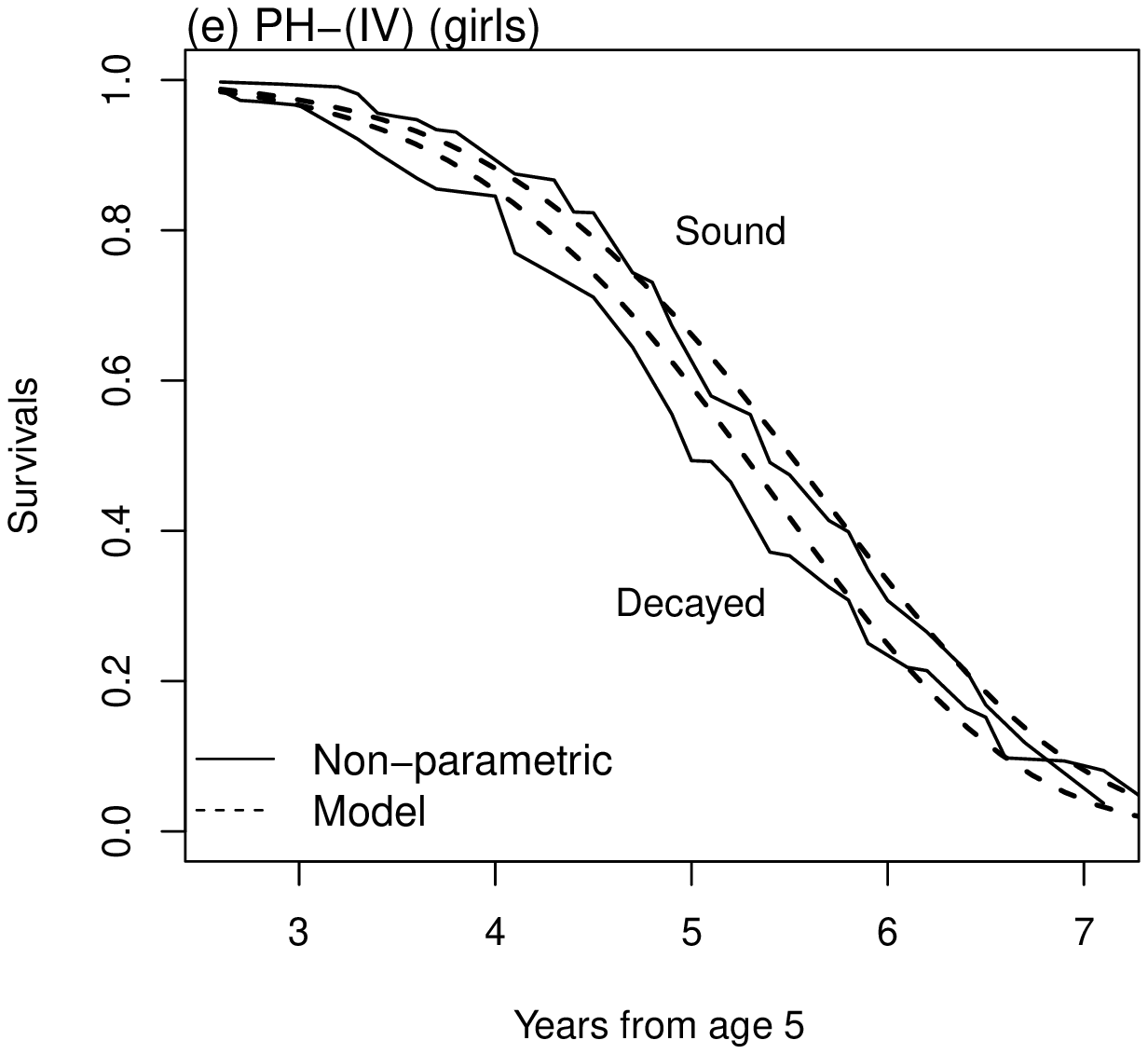}\\[0.1cm]
			\qquad{\bf (c)}\quad MPR(IV) -- Boys & \qquad{\bf (d)}\quad MPR(IV) -- Girls \\
			\includegraphics[width=0.5\textwidth, trim = {0cm 0cm 0cm 2.09cm}, clip]{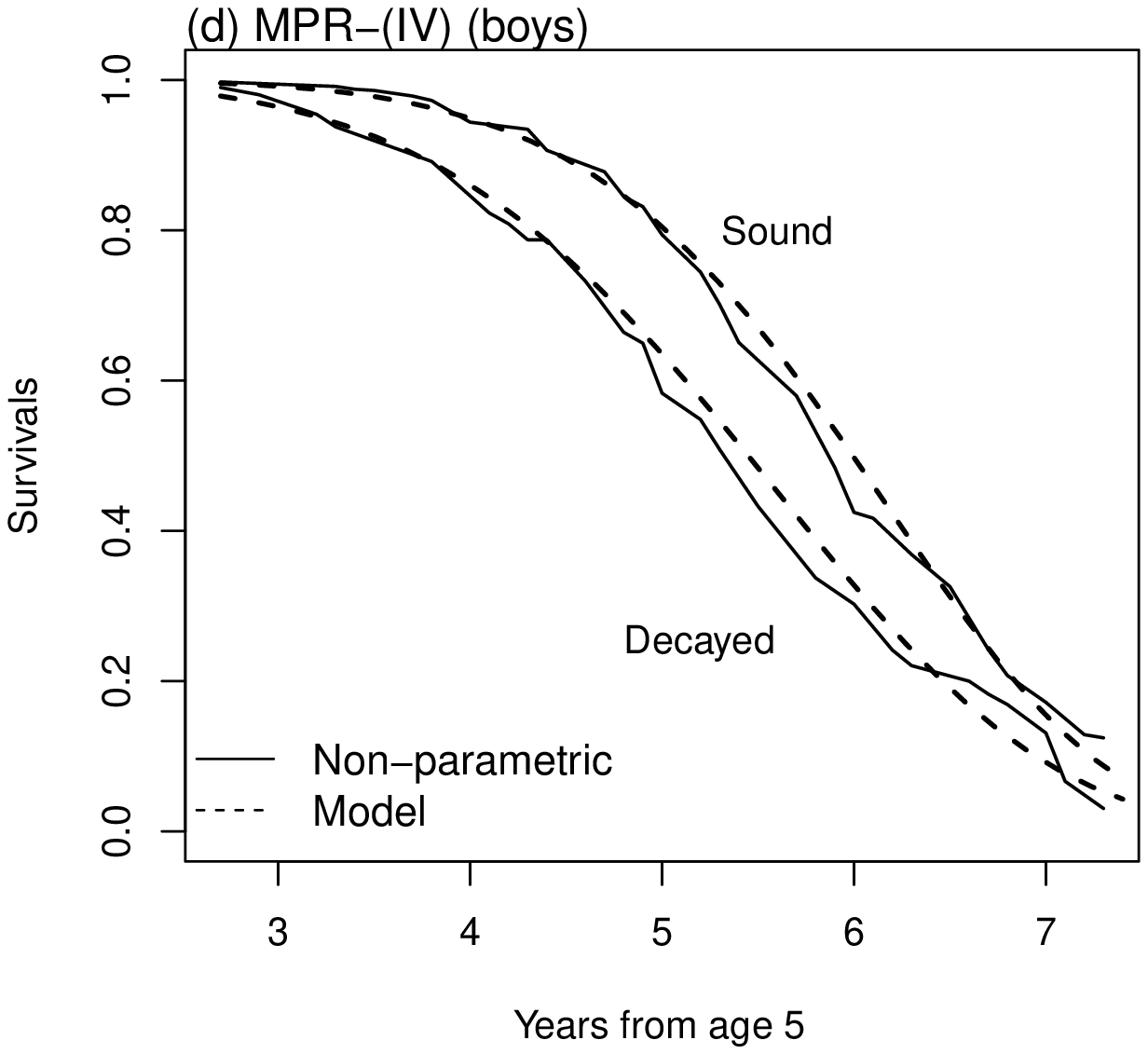} & \includegraphics[width=0.5\textwidth, trim = {0cm 0cm 0cm 2.09cm}, clip]{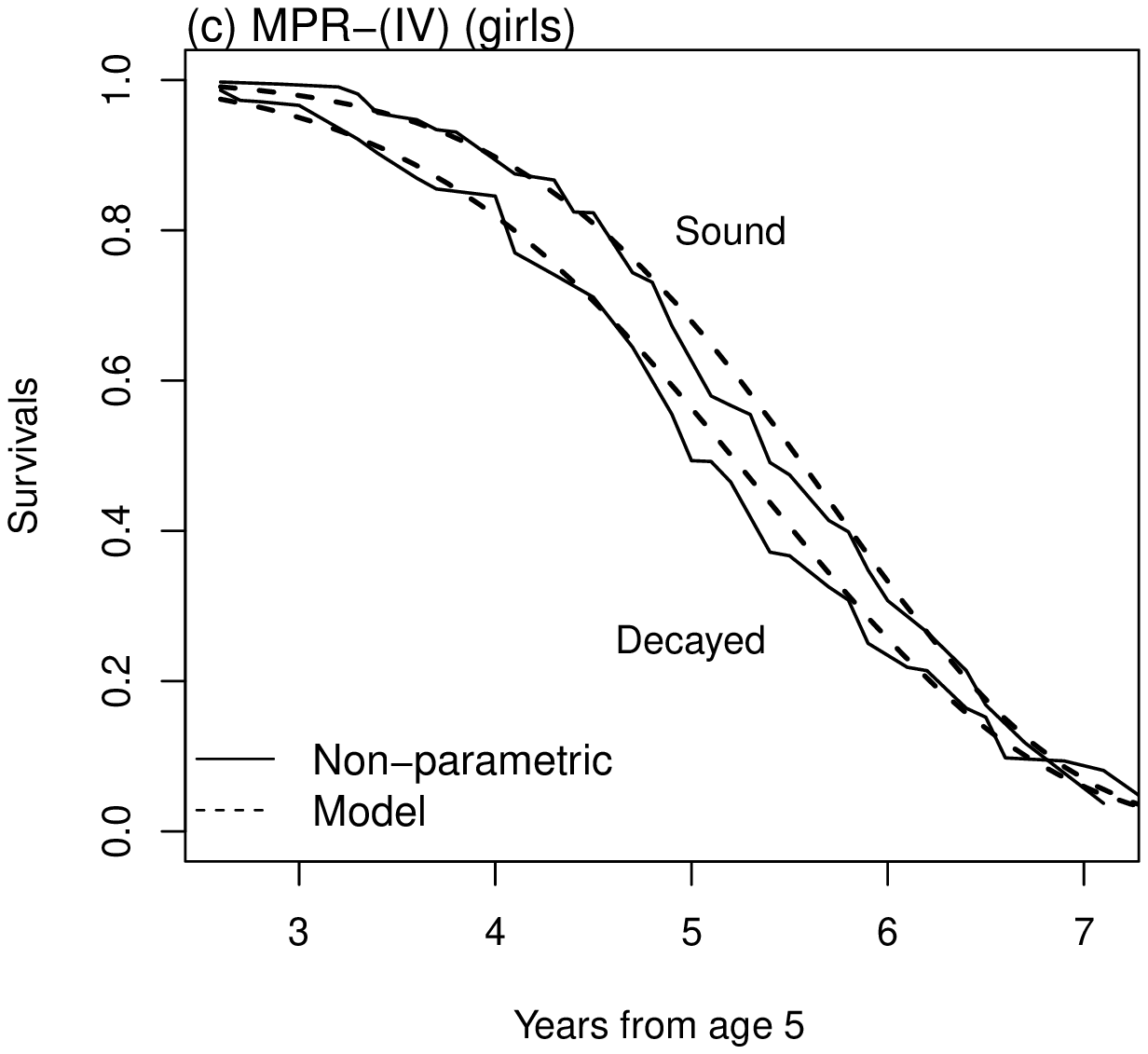}\\[0.1cm]
			\qquad{\bf (e)}\quad MPRF(IV)R -- Boys & \qquad{\bf (f)}\quad MPRF(IV)R -- Girls \\
			\includegraphics[width=0.5\textwidth, trim = {0cm 0cm 0cm 2.09cm}, clip]{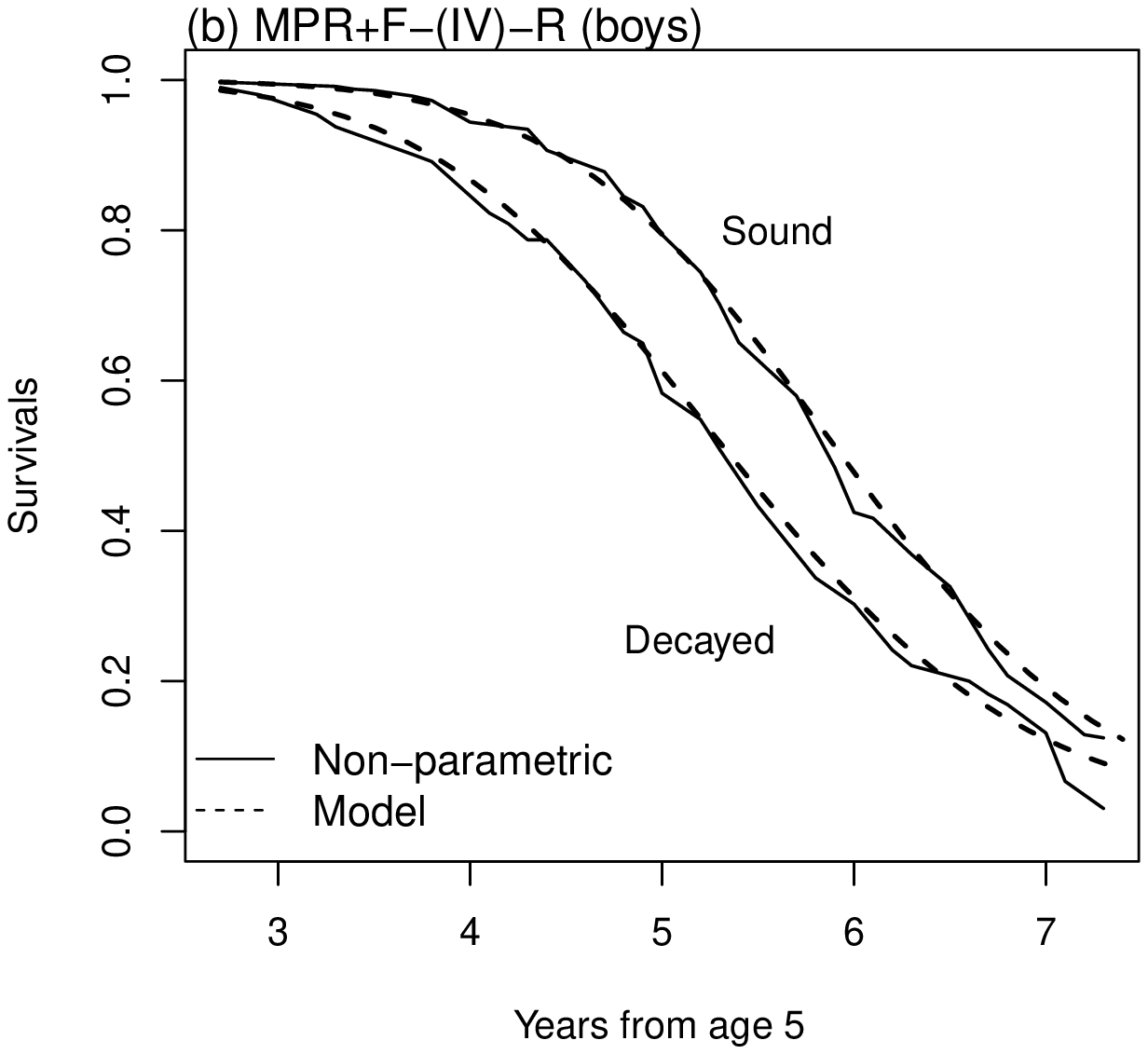} & \includegraphics[width=0.5\textwidth, trim = {0cm 0cm 0cm 2.09cm}, clip]{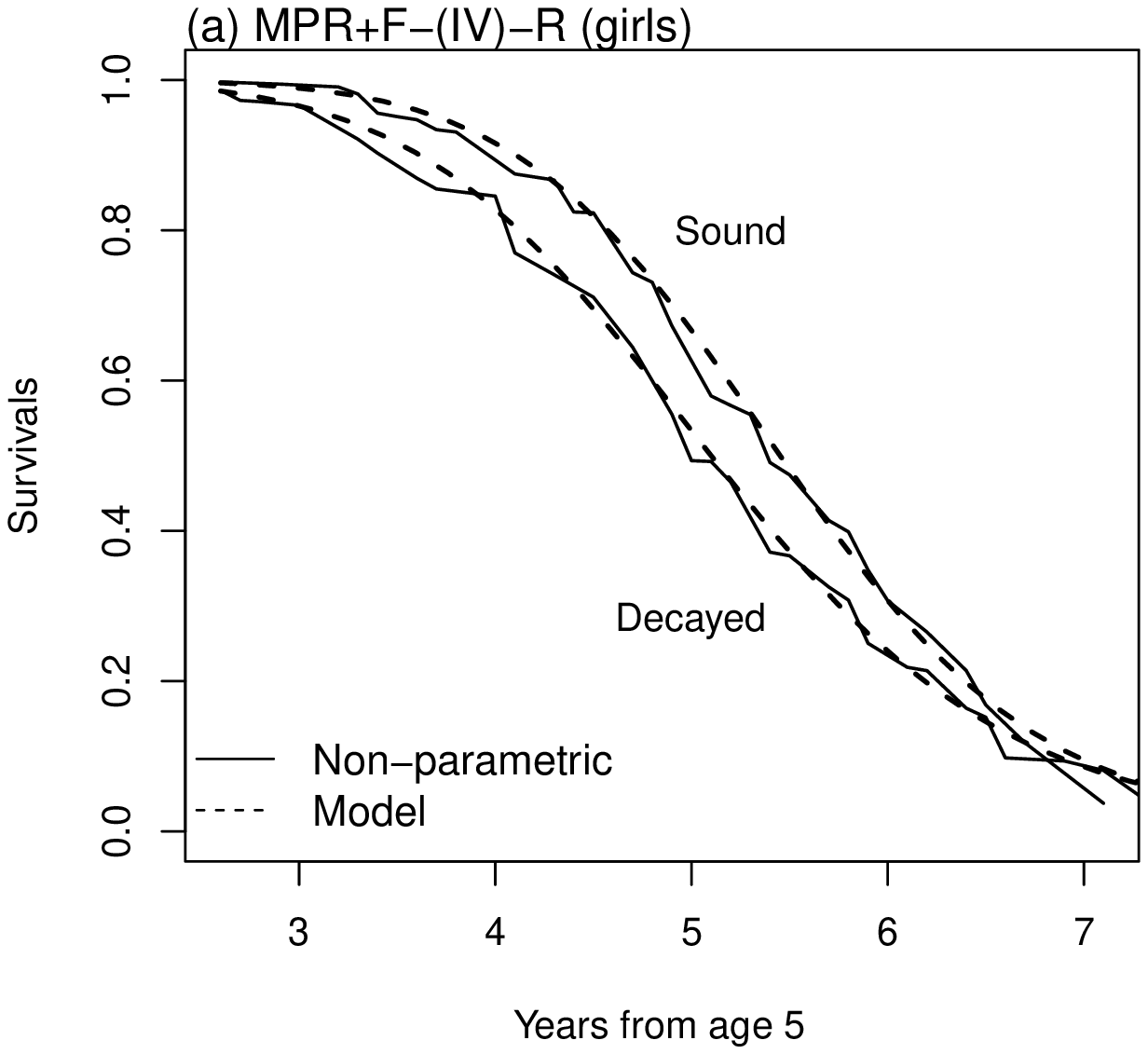}\\[-0.5cm]
		\end{tabular}
		\captionsetup{width=1.0\textwidth}
		\caption{ \small Non-parametric Turnbull survivor curves (solid) with model-based curves (dash) overlayed for PH(IV) (panels (a) and (b)),  MPR(IV) (panels (c) and (d), and  MPRF(IV)R (panels (e) and (f)). The curves for boys are on the left, while the curves for boys are to the right. Within each plot we show the sound and decayed predecessor tooth groups (i.e., non-dmf and dmf respectively). \label{fig:fits} }
	\end{center}
\end{figure}

We now discuss the merits of the various model types. First we notice in Table \ref{tab:fit} that generally the MPR models outperform their simpler PH counterparts in terms of AIC and BIC (the only exceptions being some MPRDM versus PHDM comparisons). Visually, this improvement in model fit is clear by comparing Figures \ref{fig:fits} (a) and (b) to Figures \ref{fig:fits} (c) and (d). This shows the additional value gained by modelling the shape, and highlights that one or more of the covariates have non-PH effects.
\par
The addition of frailty to PH or MPR models improves the fit in all cases. This suggests that additional heterogeneity (e.g., via unobserved covariates) exists within the data. That this is so for the MPR models is noteworthy, since the MPR model already explains variation beyond the PH model by means of its person-specific shape regression -- but, it is clear, that there is additional heterogeneity present in these data.

\par
It is noteworthy that the inclusion of  dispersion models (DM) is supported for the PH model cases, but not for the MPR cases. From (\ref{eq:margS}), we see that $\phi$ plays a role in describing the shape of the marginal distribution (albeit that $\gamma$ is the primary shape parameter). Thus, the support for DM on top of a basic PH model essentially highlights the benefit of modelling shape in addition to scale as advocated by \citet{kevin:2017}. In particular, the comparison of PHDM against MPRF allows us to assess the value of modelling shape via the frailty variance, $\phi$, or the shape parameter, $\gamma$. The latter models outperform the former in terms of AIC and BIC suggesting that there is more value in modelling the shape than the frailty variance, at least, in these data. This is perhaps not surprising since $\gamma$ is the main shape parameter as mentioned.
\par
Overall, we find that the best model in terms of AIC is MPRF(IV). This is the MPRF model with $\lambda$ and $\gamma$ regression components both containing sex, dmf, and the sex$\times$dmf interaction. The best model in terms of BIC is the MPRF(III) model, obtained by omitting the interaction term from the regression components of MPRF(IV).  \citet{gomez:2009} did not consider a non-PH process for \emph{time to emergence} of tooth24. Here, both of these best-fitting models are non-PH, and the non-PH-ness arises in two ways: (a) the shape parameter, $\lambda$, depends on covariates, and, (b) through the presence of frailty. The coefficients for these two models are shown in Table \ref{tab:red} where we see that, in MPRF(III), the sex effect appears to be statistically significant in neither the scale nor shape. A naive approach to variable selection might treat the scale and shape regression components completely separately, thereby removing sex entirely from the model. Such a removal would bring us back to MPRF(II) which, from Table \ref{tab:fit}, has much higher AIC and BIC values. This highlights the more involved nature of variable selection within MPR models, i.e., we cannot simply rely on separate Wald tests for (potentially highly) correlated scale and shape effects (see \citet{kevin:2017} for further details). In this case, it is clear that the sex effect is required, but eliminating it from the shape yielded model MPRF(III)R which has improved AIC and BIC; note that eliminating sex from the scale instead produces similar AIC and BIC values (not shown). Similarly, careful reduction of MPRF(IV) yielded MPRF(IV)R which has the lowest AIC and BIC of all models considered. Both of these reduced models are shown in Table \ref{tab:red} and, of course, the reduction produces more interpretable models.
\par
MPRF(IV)R is the best-fitting model by both AIC and BIC measures, and we can confirm that the fit is near-perfect via Figures \ref{fig:fits} (e) and (f). Before we interpret this model however, we first highlight PH(IV) shown in Table \ref{tab:red} which is the simplest full-covariate model fitted. The scale coefficients of this model suggest that tooth24 emerges first in girls with dmf, then girls without dmf and boys with dmf (whose emergence times are close), and, lastly, boys without dmf. On the other hand, model MPR(IV), also shown in Table \ref{tab:red}, directly extends this model and improves the fit which suggests that in fact something more complex arises. Returning to MPRF(IV)R, because this has a reduced shape covariate structure (compared with MPR(IV)), we can see that dmf has a highly non-PH effect due to its appearance in the shape component.

\begin{table}[!htbp]
	\footnotesize\centering
			\caption{Selection of fitted models including additional reduced models\label{tab:red}}
			\begin{tabular}{c@{\qquad\quad}c@{~~~}c@{~~~}c@{}c@{\qquad\quad}c@{~~~}c@{~~~}c@{}c}
					\toprule
				&&&&&&&\\[-0.2cm]
				&  \multicolumn{3}{c}{MPRF(III)} && \multicolumn{3}{c}{MPRF(IV)}   & \\
				\cline{2-4} \cline{6-9}
				&&&&&&&\\[-0.2cm]
				&  Scale & Shape & Frailty &&  Scale & Shape & Frailty & \\
				intercept          &  -12.97\,(0.42) &  1.98\,(0.04) & -0.45\,(0.15) &&  -13.68\,(0.54) &  2.03\,(0.04) & -0.48\,(0.16)  &\\
				girl               &    0.19\,(0.30) &  0.02\,(0.03) & ---           &&    1.52$^*$(0.59) & -0.07\,(0.05) & ---            &\\
				dmf                &    2.73$^*$(0.38) & -0.19$^*$(0.03)  & ---      &&    3.36$^*$(0.57) & -0.23$^*$(0.05) & ---            &\\
				girl$\times$dmf    &   --- & ---           & ---                     &&   -1.08\,(0.76) &  0.06\,(0.06) & ---               &\\[0.3cm]
				&  \multicolumn{3}{c}{$\text{AIC} = 10957.2$\qquad $\text{BIC} = 11001.9$} &&  \multicolumn{3}{c}{$\text{AIC} = 10950.1$\qquad $\text{BIC} = 11007.6$}  &\\[0.1cm]
				&  \multicolumn{3}{c}{$\text{dAIC} = 7.1$\qquad $\text{dBIC} = 0.0$} &&  \multicolumn{3}{c}{$\text{dAIC} = 0.0$\qquad $\text{dBIC} = 5.7$}  &\\[1cm]
				&  \multicolumn{3}{c}{MPRF(III)R} && \multicolumn{3}{c}{MPRF(IV)R}   & \\
				\cline{2-4} \cline{6-9}
				&&&&&&&\\[-0.2cm]
				&  Scale & Shape & Frailty                        &&  Scale & Shape & Frailty & \\
				intercept          &  -13.05\,(0.42) &  1.99\,(0.03) & -0.46\,(0.15)   &&  -13.22\,(0.43) &  1.99\,(0.03) & -0.46\,(0.15) &\\
				girl               &    0.47$^*$(0.06) & ---           & ---           &&    0.62$^*$(0.08) & ---           & ---          &\\
				dmf                &    2.65$^*$(0.37) & -0.18$^*$(0.03) & ---         &&    2.93$^*$(0.39) & -0.19$^*$(0.03) & ---          &\\
				girl$\times$dmf    &    ---          & ---           & ---             &&   -0.33$^*$\,(0.11) & ---           & ---          &\\[0.3cm]
				&  \multicolumn{3}{c}{$\text{AIC} = 10956.1$\qquad $\text{BIC} = 10994.4$} &&  \multicolumn{3}{c}{$\text{AIC} = 10948.6$\qquad $\text{BIC} = 10993.3$}  &\\[0.1cm]
				&  \multicolumn{3}{c}{$\text{dAIC} = 6.0$\qquad $\text{dBIC} = -7.5$} &&  \multicolumn{3}{c}{$\text{dAIC} = -1.6$\qquad $\text{dBIC} = -8.7$}  &\\[1cm]
				&  \multicolumn{3}{c}{PH(IV)} && \multicolumn{3}{c}{MPR(IV)}   & \\
				\cline{2-4} \cline{6-9}
				&&&&&&&\\[-0.2cm]
				&  Scale & Shape & Frailty &&  Scale & Shape & Frailty & \\
				intercept          &  -9.95\,(0.16)    & 1.68\,(0.02) & ---            &&  -11.82\,(0.39) &  1.86\,(0.03)     & ---           &\\
				girl               &   0.43$^*$(0.05)  &    ---       & ---            &&    1.64$^*$(0.50) & -0.11$^*$(0.04) & ---           &\\
				dmf                &   0.45$^*$(0.06)  &    ---       & ---            &&    3.08$^*$(0.48) & -0.26$^*$(0.05) & ---           &\\
				girl$\times$dmf    &  -0.21$^*$(0.08)  &    ---       & ---            &&   -1.08\,(0.63) &  0.07\,(0.06)     & ---           &\\[0.3cm]
				&  \multicolumn{3}{c}{$\text{AIC} = 11050.3$\qquad $\text{BIC} = 11082.3$} &&  \multicolumn{3}{c}{$\text{AIC} = 11003.4$\qquad $\text{BIC} = 11054.4$}  &\\[0.1cm]
				&  \multicolumn{3}{c}{$\text{dAIC} = 100.2$\qquad $\text{dBIC} = 80.3$} &&  \multicolumn{3}{c}{$\text{dAIC} = 53.2$\qquad $\text{dBIC} = 52.5$}  &\\[0.1cm]
			\bottomrule
			\end{tabular}
			\vskip 2mm
			\captionsetup{width=1.00\textwidth}
			\caption*{\footnotesize
				``girl$\times$dmf'' represents the interaction between being a girl and having dmf (primary predecessor tooth decayed / missing / filled); MPRF(III)R is a reduced form of MPRF(III) and, similarly, MPRF(IV)R is a reduced form of MPRF(IV); AIC, BIC, dAIC and dBIC are as described in Table \ref{tab:fit}. Note that regression coefficients with asterisk are statistically significant at 5\% level according to the Wald test (but we do not highlight statistically significant intercepts).}
\end{table}

\begin{figure}[!tbp]
	\begin{center}
		\begin{tabular}{@{\hspace{-0.2cm}}l@{\quad}l}
			\qquad{\bf (a)}\quad PH(IV) Hazard Functions & \qquad{\bf (b)}\quad MPRF(IV)R Hazard Functions \\
			\includegraphics[width=0.5\textwidth, trim = {0cm 0cm 0cm 2cm}, clip]{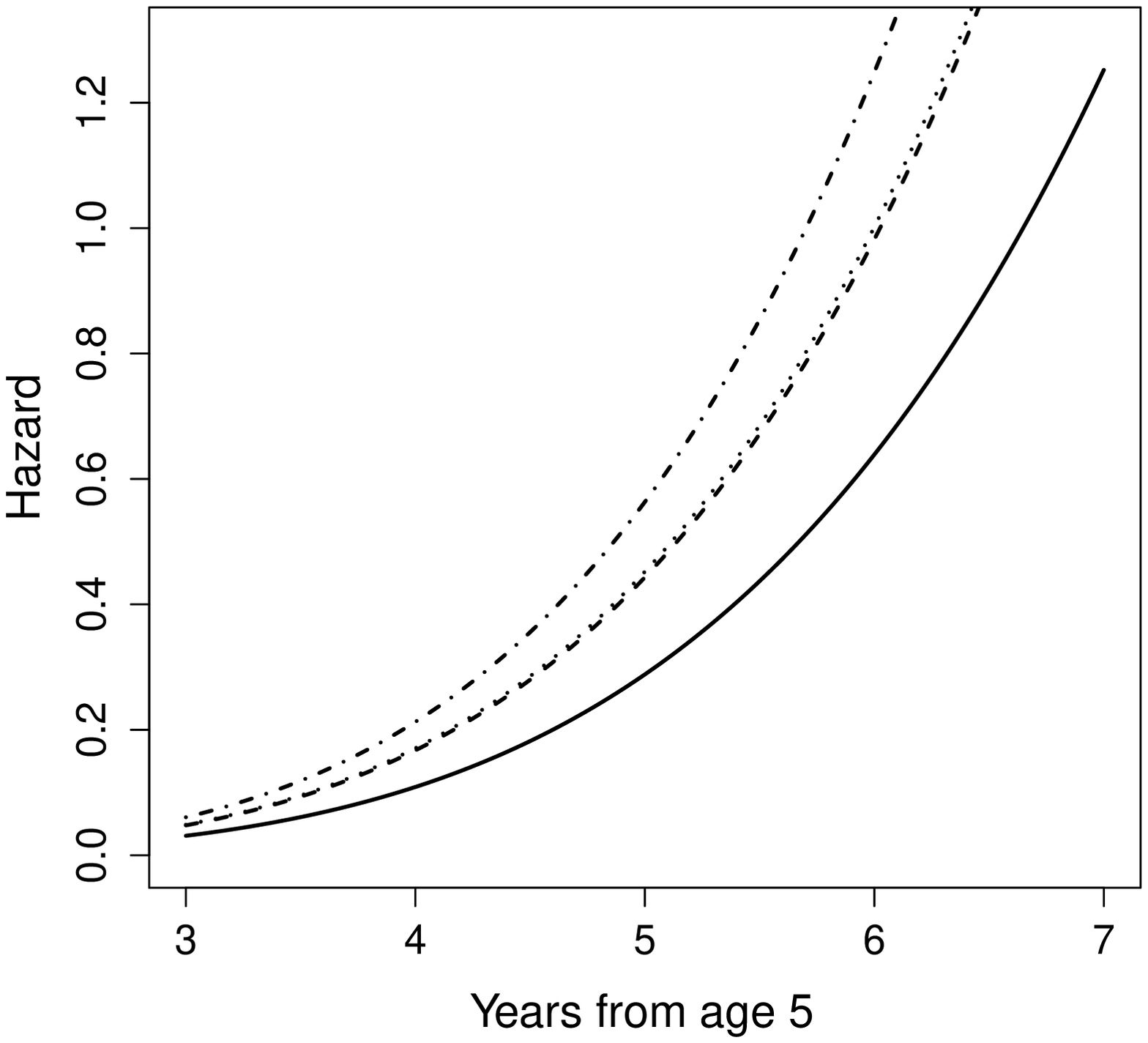} & \includegraphics[width=0.5\textwidth, trim = {0cm 0cm 0cm 2cm}, clip]{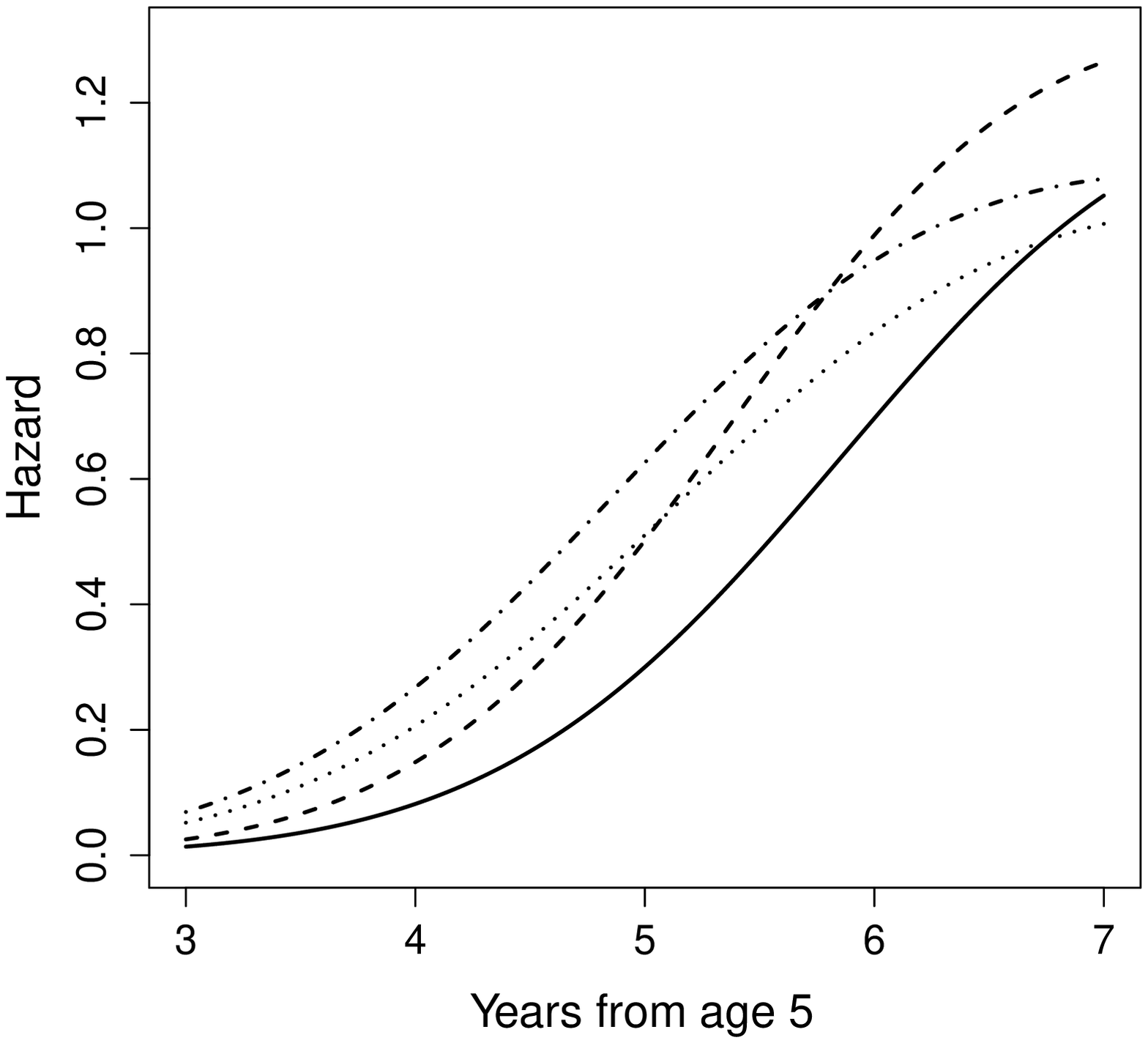}\\[0.3cm]
			\qquad{\bf (c)}\quad PH(IV) Hazard Ratios & \qquad{\bf (d)}\quad MPRF(IV)R Hazard Ratios \\
			\includegraphics[width=0.5\textwidth, trim = {0cm 0cm 0cm 2cm}, clip]{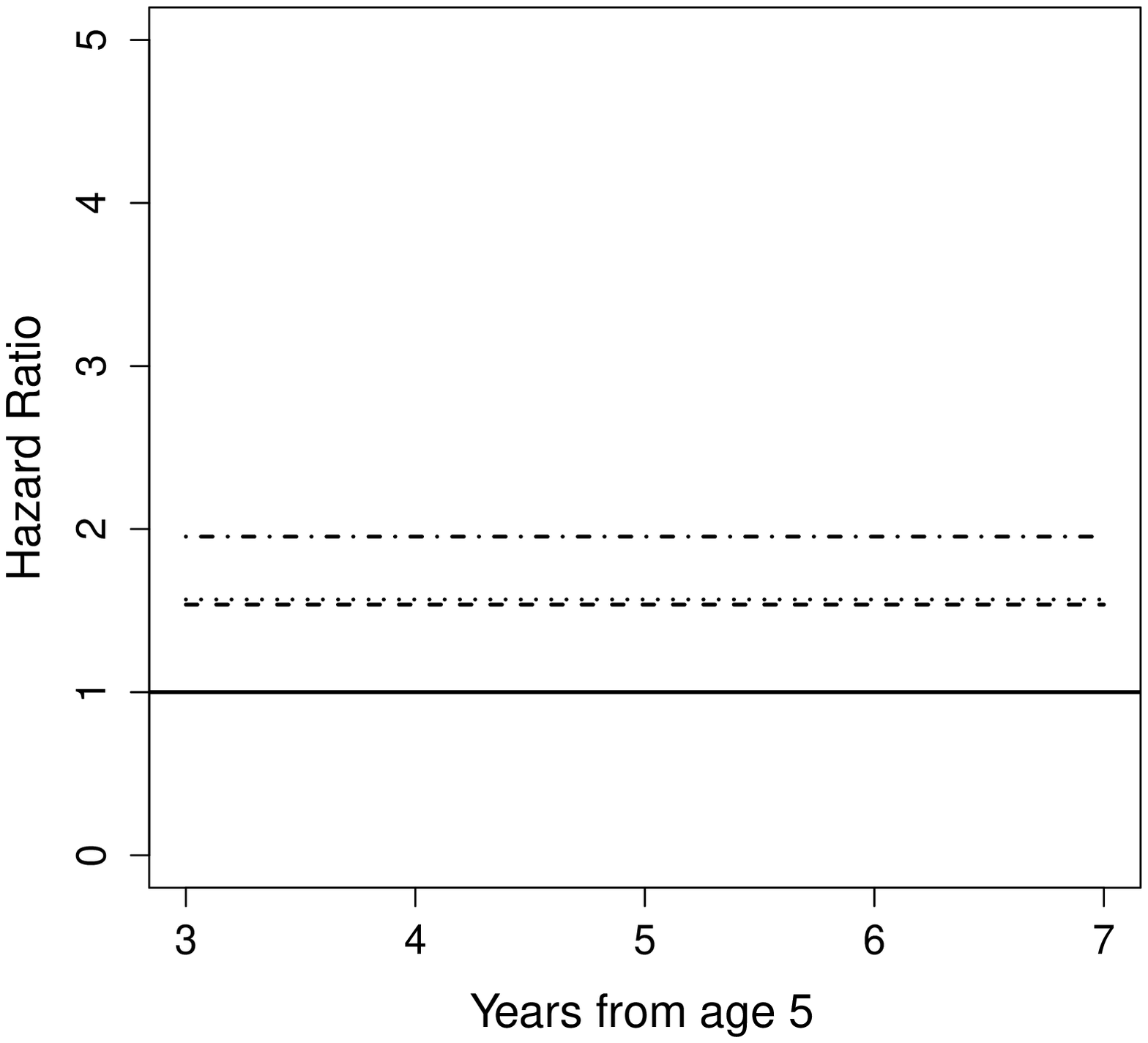} & \includegraphics[width=0.5\textwidth, trim = {0cm 0cm 0cm 2cm}, clip]{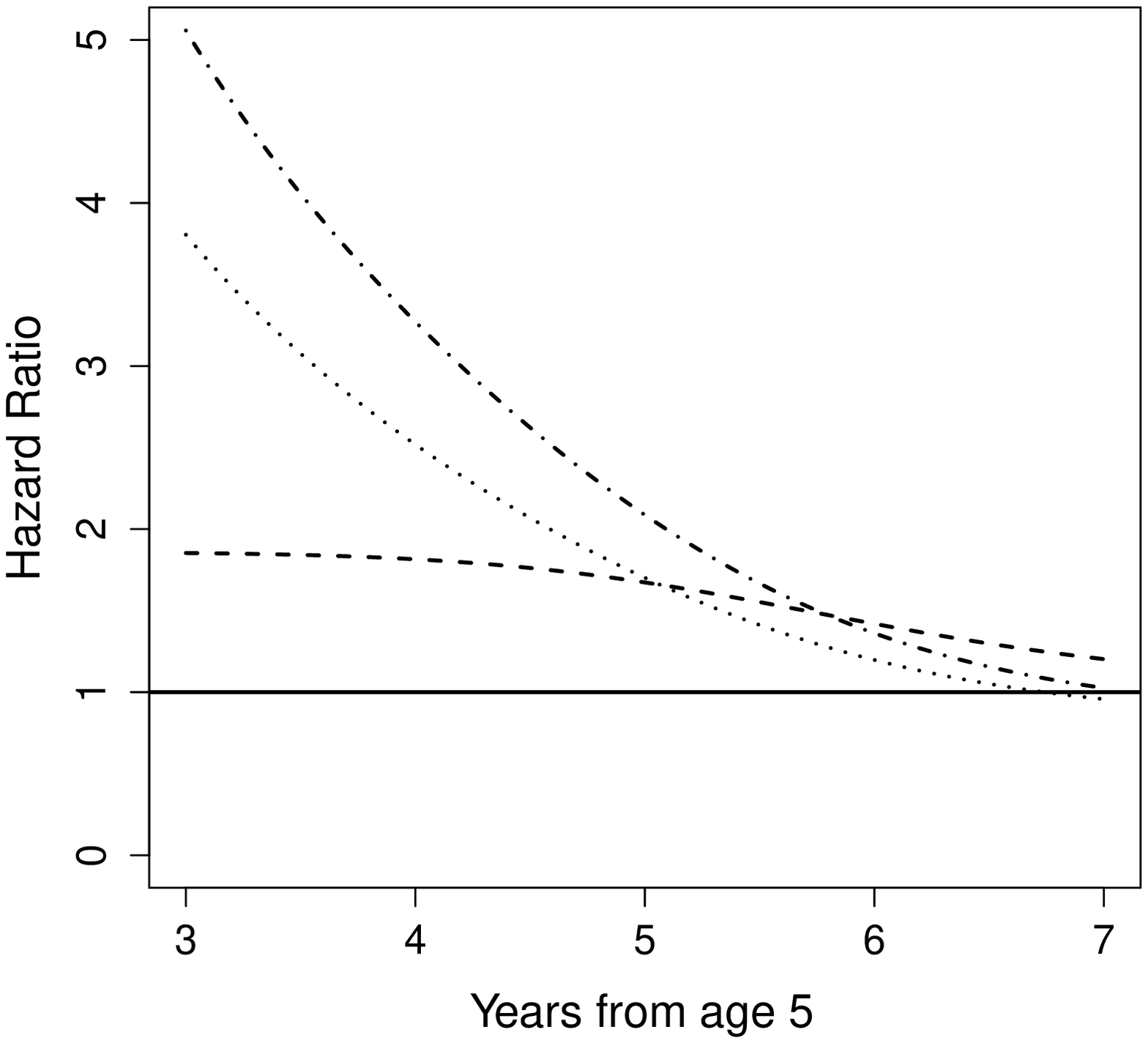}\\[0.3cm]
			\multicolumn{2}{c}{\includegraphics[width=0.7\textwidth, trim = {2.2cm 6cm 1.15cm 5.5cm}, clip]{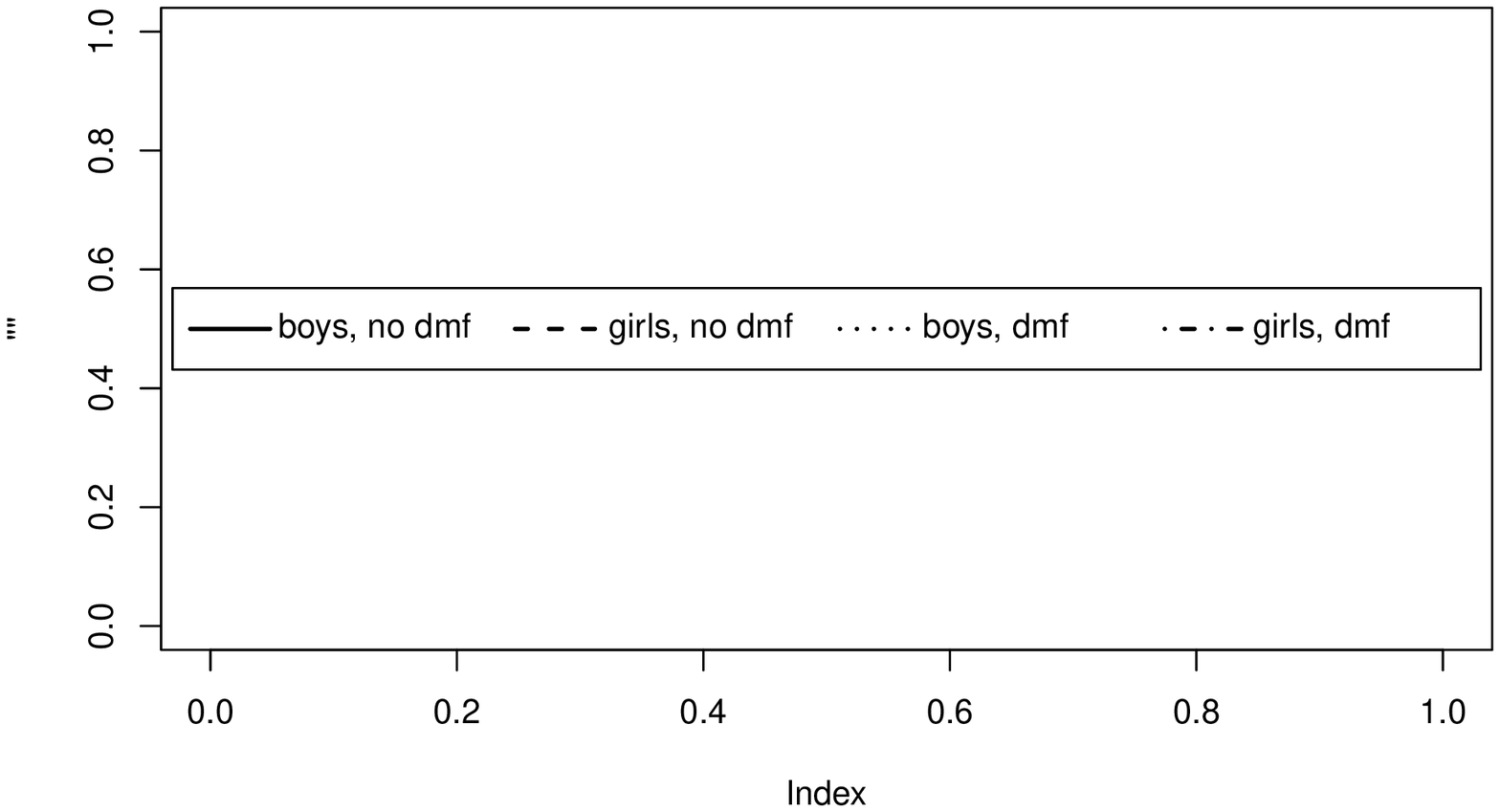}}
		\end{tabular}
		\captionsetup{width=1.0\textwidth}
		\caption{ \small Hazard functions (panels (a) and (b)) and hazard ratios (panels (c) and (d)) for boys without dmf (solid), girls without dmf (dash), boys with dmf (dot), and girls with dmf (dash-dot) for the PH(IV) (panels (a) and (c)) and MPR(IV)R (panels (b) and (d)) models respectively. Note that, for the hazard ratios, we are using boys without dmf (the lowest hazard group) as the reference group for the four groups formed via the sex$\times$dmf interaction.\label{fig:hthr}}
	\end{center}
\end{figure}

Figures \ref{fig:hthr} (a) and (b) show the hazard functions for the four groups defined by the sex$\times$dmf interaction for PH(IV) and MPRF(IV)R, while Figures \ref{fig:hthr} (c) and (d) display hazard ratios relative to the boys without dmf group.  We can see that the MPRF(IV)R model permits quite different hazard shapes for each group, whereas, for the PH(IV) model the shapes are constrained to be the same; in both cases the hazards increase with time, indicating the inevitability of the emergence of tooth24 later in time. Overall, both models agree in terms of the highest and lowest emergence hazards which correspond, respectively, to girls with dmf and boys without dmf groups. On the other hand, within the MPRF(IV)R model, it seems that boys with dmf have a higher hazard than girls without dmf earlier in time which reverses later in time -- at about time point 5 (i.e., when the children are aged 10); this crossing effect cannot be handled by the simpler PH(IV) model which forces these two groups to be equal. Table \ref{tab:med} shows the estimated median emergence times for the four groups under the PH(IV) and MPRF(IV)R models; the results are in line with the hazard-based interpretations.

\begin{table}[!htbp]
	\small\centering
		\caption{Estimated median emergence times with 95\% confidence interval\label{tab:med}}
		\begin{tabular}{cc@{\qquad}c@{\qquad}c}
			\toprule
			Dmf & Sex   &  PH(IV) & MPRF(IV)R \\
			\hline
			Yes & Girls &   5.27~~[5.19  5.34] & 5.10~~[5.01  5.20] \\
			Yes & Boys  &   5.49~~[5.40  5.57] & 5.35~~[5.26  5.44] \\
			No  & Girls &   5.51~~[5.43  5.58] & 5.49~~[5.41  5.57] \\
			No  & Boys  &   5.97~~[5.88  6.05] & 5.98~~[5.89  6.06] \\
				\bottomrule
		\end{tabular}
		\vskip 2mm
		\captionsetup{width=1.0\textwidth}
		\caption*{\footnotesize Recall that we have modelled $\text{age}-5$ years so that adding 5 to the above emergence times gives the emergence age.}
\end{table}

%
%
%

\begin{figure}[htbp]
	\begin{center}
		\begin{tabular}{@{\hspace{-0.2cm}}l@{\quad}l}
			\qquad{\bf (a)}\quad Dmf hazard ratio in boys & \qquad{\bf (b)}\quad Dmf hazard ratio in girls \\
			\includegraphics[width=0.5\textwidth, trim = {0cm 0cm 0cm 2cm}, clip]{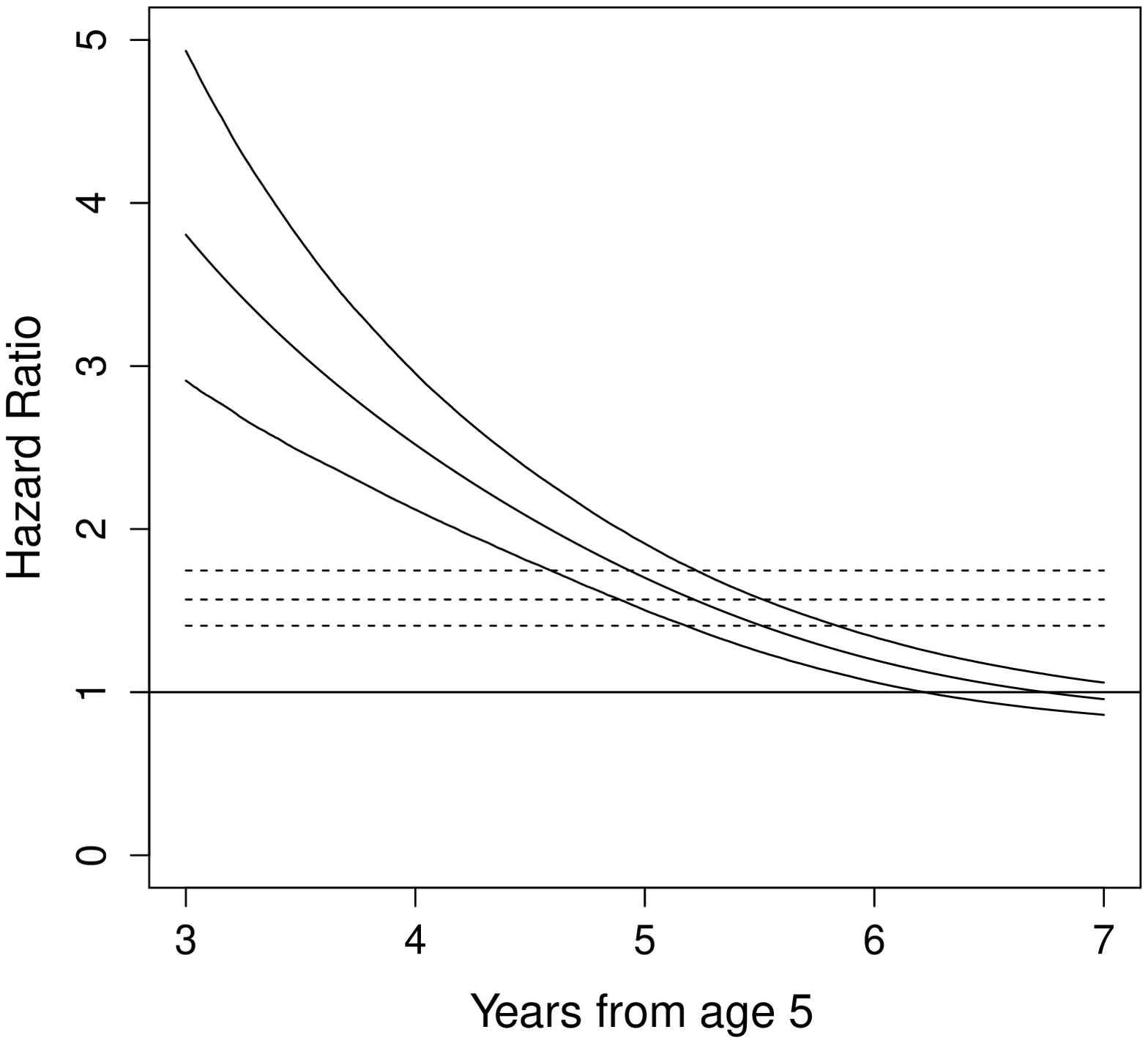} & \includegraphics[width=0.5\textwidth, trim = {0cm 0cm 0cm 2cm}, clip]{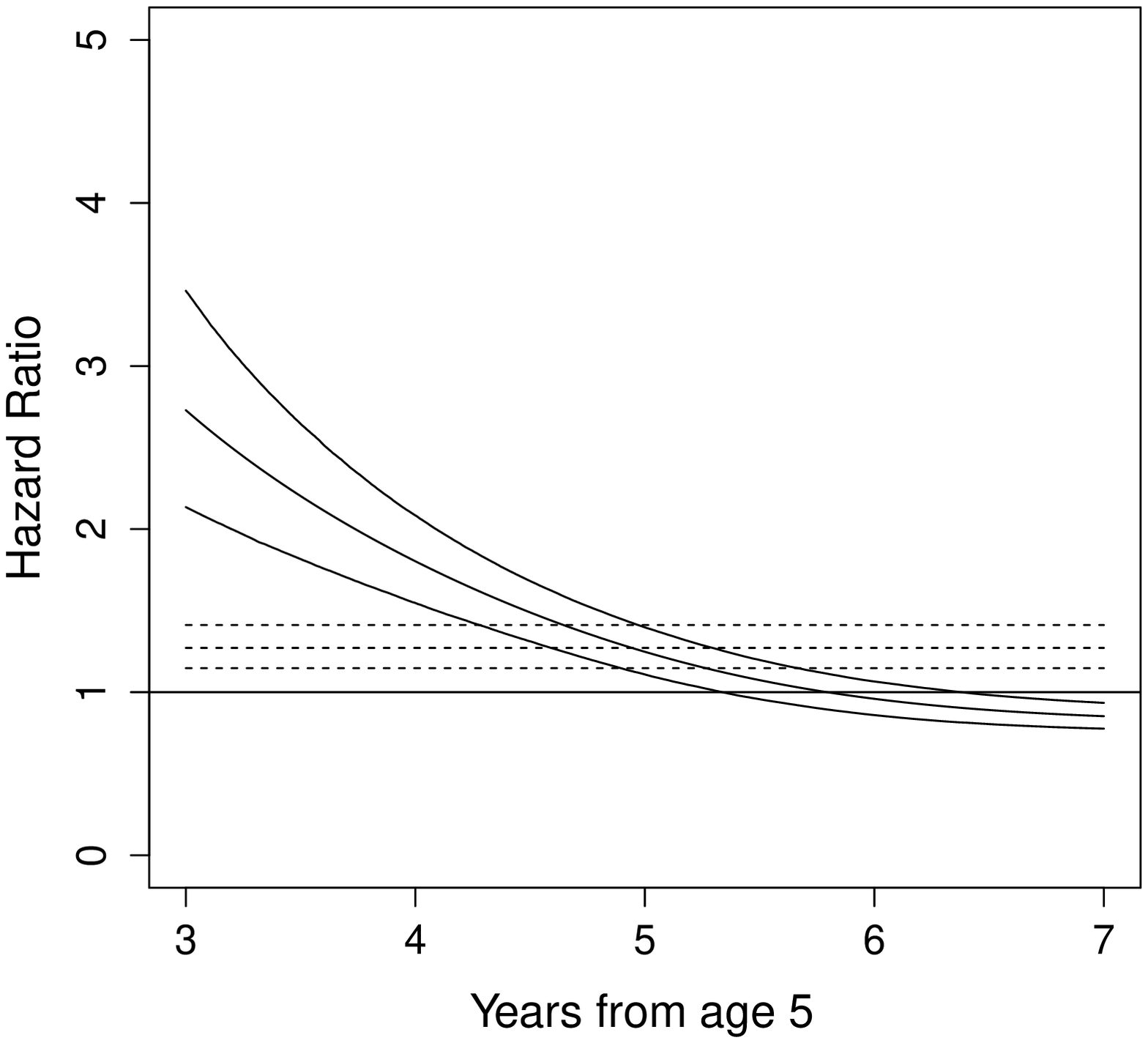}\\[0.3cm]
			\qquad{\bf (c)}\quad Sex hazard ratio in non-dmf & \qquad{\bf (d)}\quad Sex hazard ratio in dmf \\
			\includegraphics[width=0.5\textwidth, trim = {0cm 0cm 0cm 2cm}, clip]{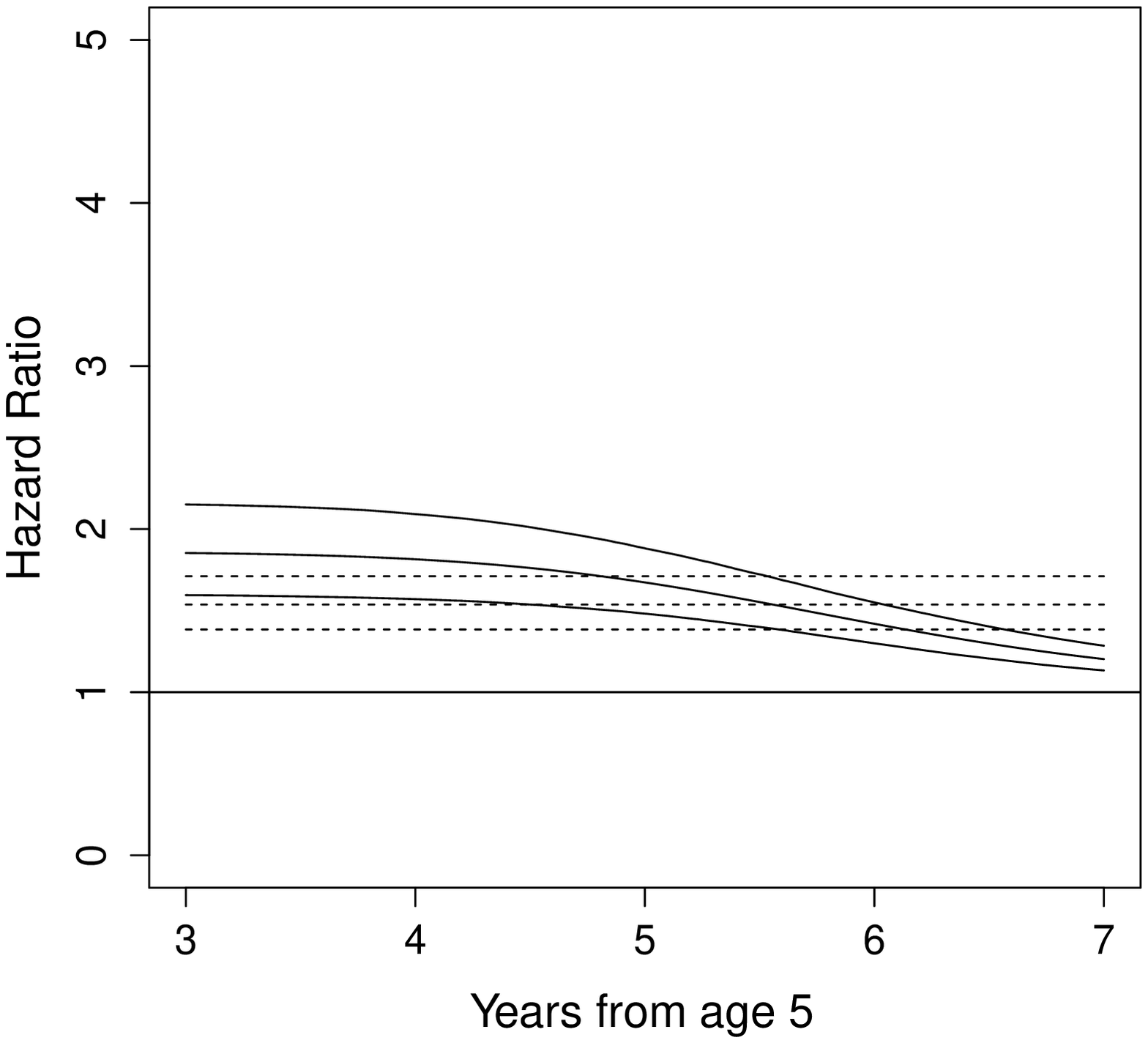} & \includegraphics[width=0.5\textwidth, trim = {0cm 0cm 0cm 2cm}, clip]{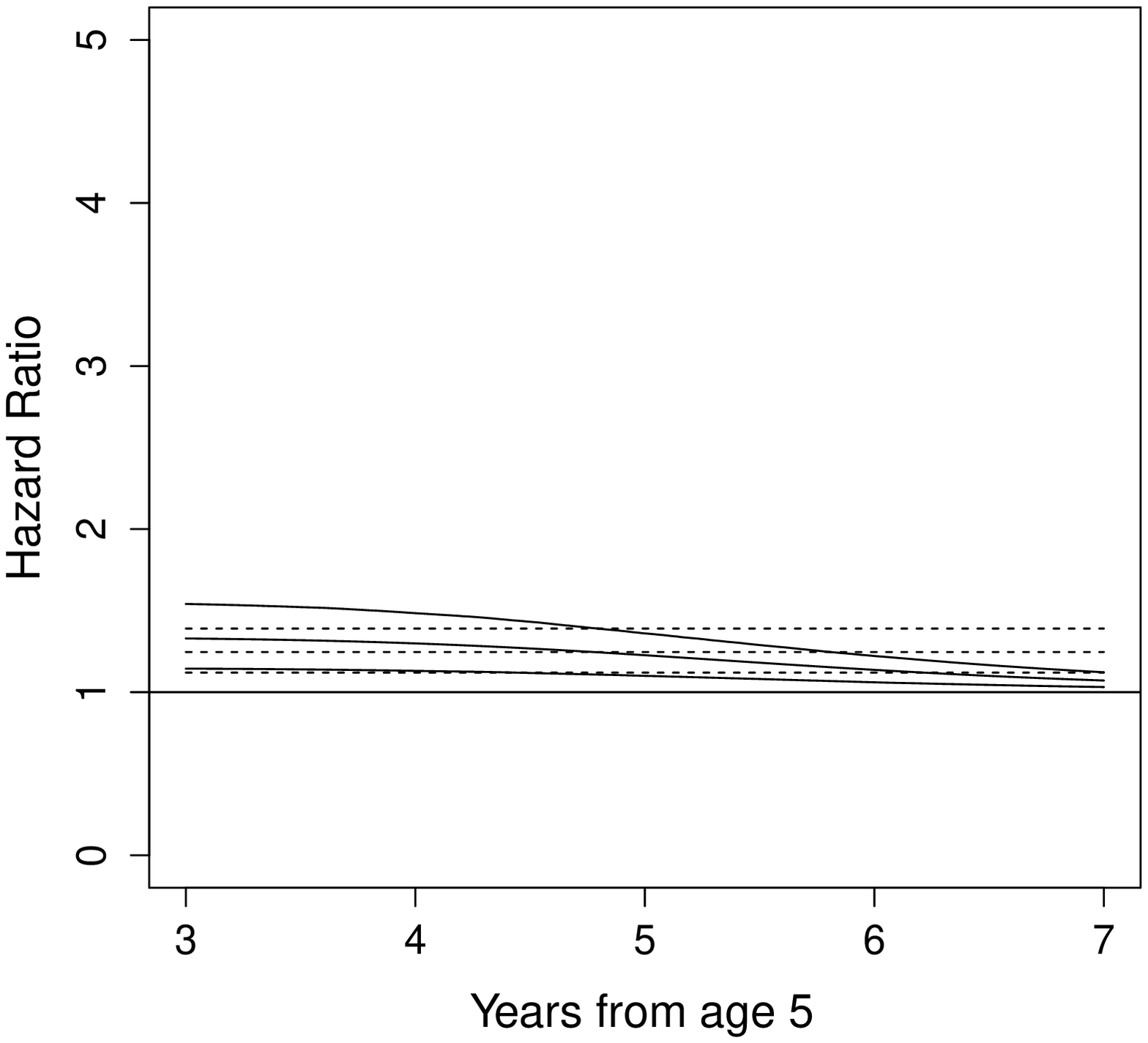}\\[0.3cm]
			\multicolumn{2}{c}{\includegraphics[width=0.7\textwidth, trim = {2.2cm 6cm 1.15cm 5.5cm}, clip]{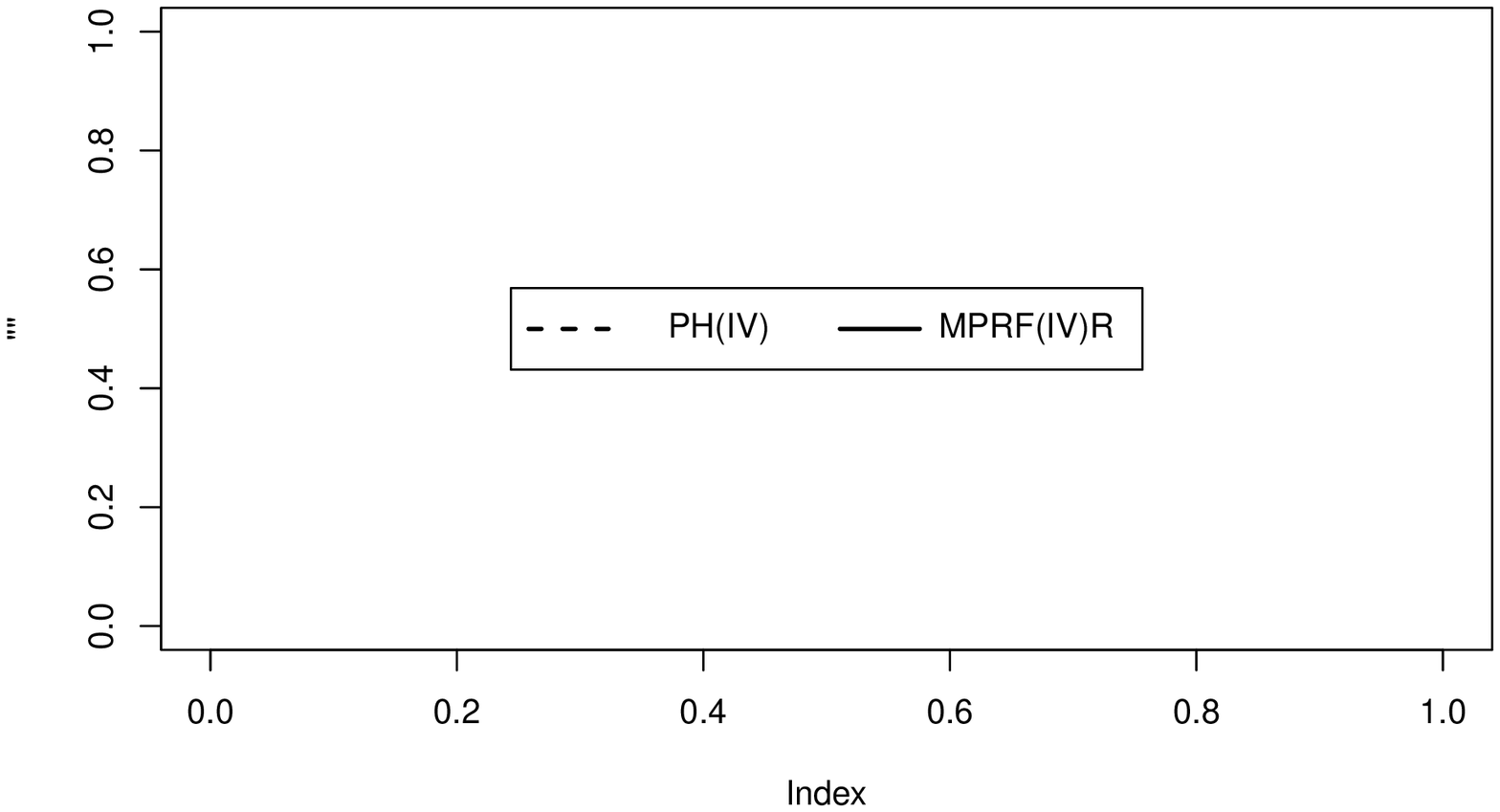}}
		\end{tabular}
		\captionsetup{width=1.0\textwidth}
		\caption{ \small Ratio of dmf to non-dmf marginal hazards in boys (panel (a)) and girls (panel (b)), and ratio of girls to boys marginal hazards in non-dmf (panel (c)) and dmf (panel (d)) groups. These marginal hazard ratios are shown for both the PH(IV) (dash) and MPR(IV)R (solid) models; a reference line at one is also shown. \label{fig:hrs}}
	\end{center}
\end{figure}

While the nature of the covariate effects can be determined by examining Figure \ref{fig:hthr} it is instructive to consider the effects of dmf and sex separately (albeit they do interact). Thus, Figures \ref{fig:hrs} (a) and (b) present the dmf hazard ratios (i.e., dmf versus non-dmf) in boys and girls. We can see that the effect of dmf, which is highly time-dependent and greater in boys, is to increase the hazard of emergence -- although, later in time, the strength of this effect reduces. Figures \ref{fig:hrs} (c) and (d)  present the sex hazard ratios (i.e., girls versus boys) in non-dmf and dmf groups. We can see that the girls have a greater hazard of emergence than boys; the effect is reduced when dmf is present. Compared with the dmf hazard ratios, we can see that the sex effect is weaker (as was apparent from Table \ref{tab:fit}) and is much less time-dependent due to the lack sex in the shape regression (in fact the MPRF(IV)R sex hazard ratios are much closer to their PH(IV) counterparts).

\section{Discussion}


In this paper we investigated the utility of MPR models in the context of interval censored data, by way of simulation and a practical application, which, to the best of our knowledge, is the first time such an investigation has been carried out. In particular, we have found that the parameters can be estimated with reasonable precision even in relatively small samples of interval censored data (albeit the most complicated model structures work best in larger samples). Moreover, we found that the MPR extension, and the additional extensions of frailty and dispersion modelling, to be fruitful in the context of the Signal Tandmobiel$^{\circledR}$ study. Thus, as we might expect, the  utility of the MPR Weibull model, and indeed the MPR framework in general, extends beyond right-censored data and the specific lung cancer application considered in \citet{kevin:2017}
\par
Our analysis of the tooth data considered a variety of additional model structures not previously explored in the existing literature \citep{bogaerts:2002, gomez:2009, kl:2005, kl:2009}. It is noteworthy that \citet{kl:2005}, who developed a spline-based AFT model, suggest that ``parametric methods do not offer enough flexibility to correctly model survival data''. In contrast, we have demonstrated the appropriateness of relatively simpler parametric models for these data which achieve flexibility through a combination of multi-parameter regression and frailty modelling. Interestingly, \citet{kl:2005} and \citet{kl:2009} briefly considered a dispersion model extension of their AFT model which they refer to as a mean-scale model (citing earlier work by \citet{panmac:2003} on mean-covariance modelling in longitudinal studies). While an investigation of MPR approaches was not their focus, they nonetheless found, like us, that an MPR extension yielded a superior fit to the data.
\par
It is fortunate that these data are sufficiently extensive to permit the investigation of such models. We find that to emergence of the permanent upper left first premolars depends on sex and dmf. In particular, emergence times are significantly earlier in children whose predecessor tooth was decayed, missing, or filled (dmf), and that emergence times are earlier in girls. However, we have also found that frailty effects are supported within the data, i.e., there may be further unmeasured features at play. The time-dependent nature of the dmf hazard ratios is quite interesting, and suggests that the dmf group becomes more like the non-dmf group later in time. Recall that the dmf group is a mixture of individuals with decayed, missing, and filled teeth. With this in mind, one might speculate that filled teeth are more similar to sound teeth (non-dmf teeth) with larger emergence times, while missing teeth are perhaps quite different with shorter emergence times. This, at least, would be a frailty interpretation of time varying effects, i.e., there is a mixture of groups, some of which ``fail'' earlier. However, the time variation is much greater than that supported by frailty alone due to the presence of the significant dmf shape effect within the MPRF(IV)R model.

\par
The extension of the MPR framework to interval-censored survival data with frailty permits the examination of a variety of potential data structures. An appealing aspect of this approach is that the breadth of models supported exist within a reasonably straightforward parametric setup which is not computationally intensive. Such a framework provides a practical and useful adjunct to existing methods which may reveal new insights.

\section*{Acknowledgements}

This research was supported by the Irish Research Council (New Foundations Award).

\newpage

\bibliographystyle{apalike}
\bibliography{refs}

\end{document}